\documentclass[journal]{IEEEtran}

\usepackage{cite}
\usepackage{amsmath}
\usepackage[british]{babel}
\usepackage{amsfonts}         
\usepackage{amsthm}           
\usepackage{bm}               
\usepackage{units}
\usepackage{multirow}            
\usepackage{bigdelim}
\usepackage{gensymb}
\usepackage{ifthen}
\usepackage{graphicx} 
\usepackage{dblfloatfix}
\usepackage{mathtools}
\graphicspath{{../figures/}{./}}
\DeclareGraphicsExtensions{.pdf,.jpeg,.png}
\usepackage[caption=false]{subfig}
\usepackage[OT1]{fontenc}      

\usepackage[framemethod=tikz]{mdframed}
\usepackage{pdfpages}
\usepackage{pdflscape}

\usepackage{fancyhdr}            							
\usepackage[subfigure,titles]{tocloft}      
\usepackage{amssymb}
\usepackage{tabularx}
\usepackage[inline]{enumitem}
\usepackage{soul}
\usepackage{algorithm}
\usepackage{algorithmic}
\usepackage{float}
\newfloat{algorithm}{t}{lop}

\usepackage{mdwmath}
\usepackage{mdwtab}
\usepackage{array}

\makeatletter
    \newcommand{\vast}{\bBigg@{3}}
    \newcommand{\Vast}{\bBigg@{3.5}}
    \newcommand{\vastt}{\bBigg@{4}}
    \newcommand{\Vastt}{\bBigg@{4.5}}
    \newcommand{\vAst}{\bBigg@{5}}
    \newcommand{\vAstt}{\bBigg@{5.5}}
    \newcommand{\vaSt}{\bBigg@{6}}
    \newcommand{\vaStt}{\bBigg@{6.5}}
    \newcommand{\vasT}{\bBigg@{7}}
    \newcommand{\vasTt}{\bBigg@{7.5}}
    \newcommand{\vastT}{\bBigg@{8}}
    \newcommand{\VAst}{\bBigg@{8.5}}
    \newcommand{\VAstt}{\bBigg@{9}}
    \newcommand{\VaSt}{\bBigg@{9.5}}
    \newcommand{\VaStt}{\bBigg@{10}}
    \newcommand{\VasT}{\bBigg@{10.5}}
    \newcommand{\VasTt}{\bBigg@{11}}
    \newcommand{\VastT}{\bBigg@{11.5}}
    \newcommand{\VASt}{\bBigg@{12}}
    \newcommand{\VAStt}{\bBigg@{12.5}}
    \newcommand{\VAsT}{\bBigg@{13}}
    \newcommand{\VAsTt}{\bBigg@{13.5}}
    \newcommand{\VAstT}{\bBigg@{17}}
\makeatother

\begin{document}
\bstctlcite{IEEEexample:BSTcontrol}
\title{Defining and Surveying Wireless Link Virtualization and Wireless Network Virtualization}

\author{\IEEEauthorblockN{
Jonathan van de Belt\IEEEauthorrefmark{1}, Hamed~Ahmadi\IEEEauthorrefmark{2},~and~Linda E. Doyle\IEEEauthorrefmark{1}} \\
\IEEEauthorblockA{\IEEEauthorrefmark{1}The Centre for Future Networks and Communications \-- CONNECT, Trinity College Dublin \\
Email: vandebej@tcd.ie}
\IEEEauthorblockA{\\ \IEEEauthorrefmark{2}School of Electrical and Electronic Engineering, University College Dublin, Ireland}\vspace{-2ex}}

\maketitle


\begin{abstract}

Virtualization is a topic of great interest in the area of mobile and wireless communication systems. However the term virtualization is used in an inexact manner which makes it difficult to compare and contrast work that has been carried out to date. The purpose of this paper is twofold. In the first place, the paper develops a formal theory for defining virtualization. In the second instance, this theory is used as a way of surveying a body of work in the field of wireless link virtualization, a subspace of wireless network virtualization. The formal theory provides a means for distinguishing work that should be classed as resource allocation as distinct from virtualization. It also facilitates a further classification of the representation level at which the virtualization occurs, which makes comparison of work more meaningful. The paper provides a comprehensive survey and highlights gaps in the research that make for fruitful future work.

\end{abstract}

\IEEEpeerreviewmaketitle

\section{Introduction}
\label{sect:intro}

Network Virtualization (NV) allows network services to view network resources, such as servers, routers, links, and data, in a manner that is independent from the underlying physical infrastructure, and to use these resources according to service requirements, rather than based on physical granularities \cite{Chowdhury2009}. New network functionality can be achieved using virtualization, such as providing heterogeneous networks with customizable specifications on-demand, the flexible and dynamic management of resources, new types of services, and better security and protection against equipment failure. In addition, network virtualization has the potential to enable new networking technologies and protocols to be developed much faster than they are currently, since these technologies can be tested through isolated virtual networks on existing infrastructure, while ensuring that existing services are unaffected. Lastly, network virtualization can provide cost savings and new business opportunities, through increased efficiencies and through new services and functionality.



Wireless Network Virtualization (WNV) has been proposed as an extension of (wired) network virtualization to the wireless domain, with the main difference being the wireless links. Thus most work to date has focussed on Wireless Link Virtualization (WLV). Initially, the purpose of this paper is to perform a survey of WNV, and to identify open research problems. However, the term `wireless network virtualization' carries multiple connotations. It has become an umbrella term for several differing concepts, applied at different layers of the network stack, and also to different types of network resources. In the existing literature, works such as \cite{Sachs2008, Forde2011, Wang2013Network, Wen2013Current, Liang2014, Liang2015} provide a variety of definitions for WNV, but a lack of consistency persists. A common theme of these definitions is that they regard virtualization as the abstraction and sharing/slicing of resources. As this work will emphasise, virtualization and abstraction are very different concepts, and virtualization is not necessarily limited to the sharing of resources. It is interesting to note that several authors (\!\!\cite{Nakao2010} and \cite{Wang2013Network}) have pointed out a similar vagueness and lack of clarity in the field of network virtualization.

Although several surveys on wireless network virtualization exist such as \cite{Wen2013Current, Liang2014, Wen2013, Yang2014, Richart2016}, this paper brings additional and alternative perspectives (for a more detailed description see Section \ref{subsect:lit_rev}). We develop a formal method for describing virtualization as a response to the many definitions that currently exist. We then use this formal method as a means of classifying and analysing the papers we survey which allows for a more systematic approach to the survey process.

\begin{figure}[!t]
\centering
\subfloat[]{
	\centering
	\includegraphics[height=.13\textheight]{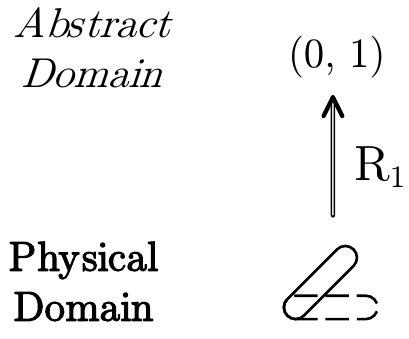}
}
\hfill
\subfloat[]{
	\centering
	\includegraphics[height=.13\textheight]{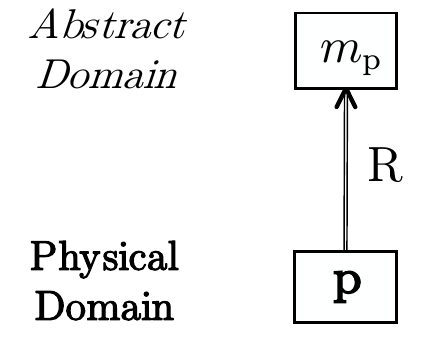}
}
\caption{Objects in the physical domain can be \textit{represented} using objects in the abstract domain, such as (a) a switch with two settings represented as a bit through $\mathrm{\fontfamily{cmr-12}\selectfont R_1}$, or (b) more generally as an object \textbf{p} represented as $m_{\textbf{p}}$ through {\fontfamily{cmr-12}\selectfont R}.}
  \label{fig:ar1}
\end{figure}

More specifically we make the following contributions:

\begin{enumerate}
\item We clarify the concepts of abstraction and representation, which are key to understanding virtualization, by drawing on a theory known as abstraction/representation theory and extending it. 
\item We propose a formal method for describing virtualization, which we call virtualization theory. 
\item We develop a test for virtualization to distinguish virtualization techniques from resource allocation techniques.
\item We survey the existing work on wireless network virtualization, and classify this work in a coherent and meaningful manner, using virtualization theory.
\item We identify several research gaps in wireless network virtualization which have not yet been addressed and propose next steps forward. 
\end{enumerate}

The paper is structured around these contributions. Section \ref{sect:abstraction} introduces key concepts such as abstraction, representation and instantiation, which are prerequisite to virtualization and important for the rest of the paper. Section \ref{sect:theory} introduces the theory on which the paper is grounded. In Section \ref{sect:resource_virtualization} we examine the constituent components of networks, and how these components can be virtualized. We give an overview of network and wireless network virtualization in Section \ref{sect:nv}, to introduce the survey. Since the main focus of this paper is on wireless link virtualization, in Section \ref{sect:survey} we perform a survey of existing literature on wireless link virtualization. This survey allows us to identify open research directions in Section \ref{sect:questions}, before concluding in Section \ref{sect:conclusion}.


\begin{figure*}[!tp]
\centering
\subfloat{
	\centering
	\includegraphics[height=.12\textheight]{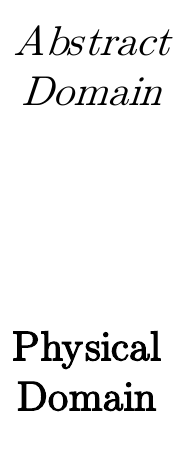}
	\addtocounter{subfigure}{-1}
}
\hfill
\subfloat[]{
	\centering
	\includegraphics[height=.12\textheight]{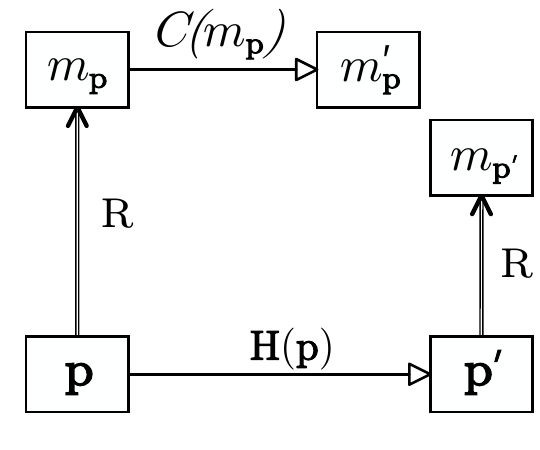}
}
\hfill
\subfloat[]{
	\centering
	\includegraphics[height=.12\textheight]{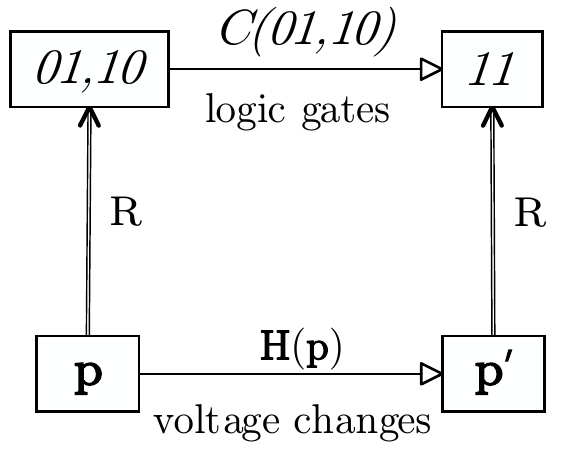}
}
\hfill
\subfloat[]{
	\centering
	\includegraphics[height=.12\textheight]{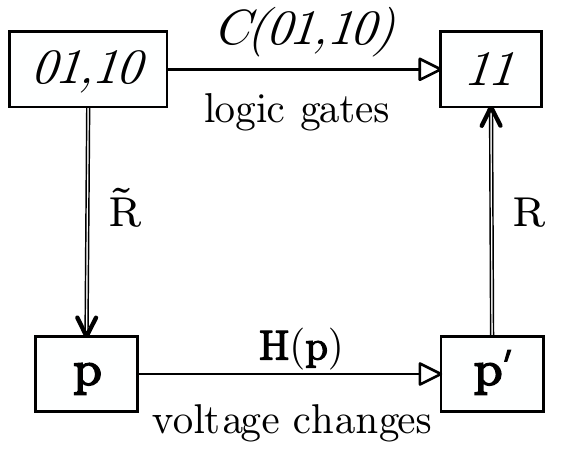}
}
\hfill
\subfloat[]{
	\centering
	\includegraphics[height=.12\textheight]{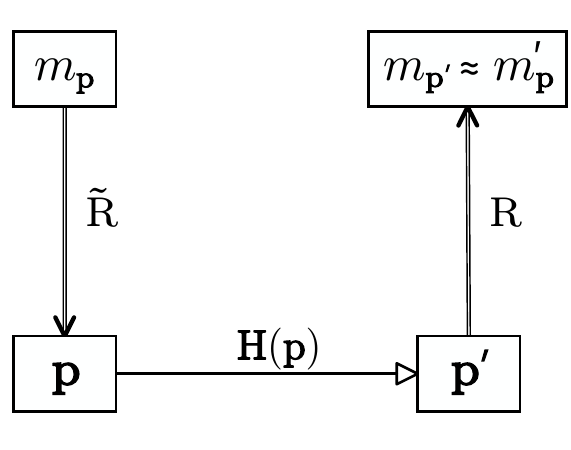}
}
\caption{(a) A physical system \textbf{p} and its representation $m_\textbf{p}$ can undergo abstract evolution, \textit{C($m_\textbf{p}$)}, or physical evolution, \textbf{H}(\textbf{p)}. (b) A commuting diagram for a binary addition, showing that an abstract evolution through logic gates, \textit{C}, is commutative with a physical evolution in voltage states, \textbf{H}. (c) A computer can evaluate an abstract mapping by instantiating the abstract object in the physical domain, performing the physical mapping, and representing it as an abstract object. (d) The most interesting use of a computer is when the abstract mapping is unknown. }
\label{fig:commute}
\end{figure*}

\section{Prerequisites to Virtualization}
\label{sect:abstraction}

There are a number of concepts, which are key to developing a formal theory of virtualization and in the opening section of the paper we carefully define these concepts. The first two of these are the concepts of abstraction and representation. Table \ref{table:terms} shows an explanation of these and other terms used in this paper.


In this paper, the term `\textit{abstraction}' means the act of ignoring or hiding details to consider general characteristics, rather than concrete realities. Thus abstraction manages the way in which systems interact, and the complexity of the interaction, by hiding details that are not relevant to the interaction. Increased abstraction allows systems to be used more easily for specific applications, but this comes at the cost of decreased flexibility and customization. Although abstraction is an important concept in computing, as it governs the interaction between humans and computers, it is not necessarily of importance to virtualization. However, it is important to note the difference between abstraction and the adjective `abstract'. The term `abstract' refers to ideas and concepts that do not have physical existence. 

The term `\textit{representation}' means to describe or symbolize something in a particular way. The term `\textit{instantiation}' means the implementation or realization of a concept or idea. Representation and instantiation are very important to computing as they describe the relationship between abstract entities and the physical world. As this paper will show later, representation and instantiation are of great importance to virtualization, since virtualization can only be done in the abstract domain, while network resources exist in the physical domain. 

We now turn to the recently developed Abstraction/Representation (AR) theory, to provide us with a formal framework of abstraction and representation \cite{Horsman2014}, \cite{Horsman2014_2}. Because this theory is very new, we explain the most important aspects here, borrowed from \cite{Horsman2014}, before we go on to extend the ideas for the purposes of this paper.

\begin{table}[h]
		\caption{Explanation of terms used in this paper}
	   	\label{table:terms}
	    \centering
	   	\begin{tabular}{l l}
	 	\hline \\ [-2ex]
	 	\textbf{Term} & \textbf{Explanation} \\ [0.5ex]
	 	\hline \\ [-2ex]
		Abstraction & \begin{tabular}{@{}l@{}} The act of ignoring details to consider general \\ characteristics \end{tabular}	\\ [2ex]
		Abstract & \begin{tabular}{@{}l@{}} Existing as a thought, idea, or concept, but without \\ physical existence \end{tabular} \\ [2ex]
		Representation & \begin{tabular}{@{}l@{}} The act of symbolising or portraying something in a \\ particular way \end{tabular} \\ [2ex]
		Instantiation & \begin{tabular}{@{}l@{}} Creating a concrete realization of a concept or idea \end{tabular} \\ [0.5ex]
		\hline \\ [-2ex]		
		\end{tabular}
\end{table}  

 
\subsection{Abstraction/Representation Theory \protect\footnote{Based on \cite{Horsman2014}}}

Abstraction/Representation theory \protect\footnote{AR Theory might be better named as Representation Theory, since it deals mostly with representation, but that name has already been taken.} is concerned with the physical domain and the abstract domain (also known as the logical domain), and the relationship between these domains. The physical domain, \textbf{P}, is defined as consisting of all physical objects, \textbf{p} $\in$ \textbf{P}. The abstract domain, M, consists of all abstract objects, $m \in$ M. For instance, a \textit{computer} is an object in the physical domain, which can be in different physical states, while a \textit{computation} is a set of objects and relations in the abstract domain. Bold font is used to indicate an object in the physical domain; italic font for an object in the abstract domain. 

\subsubsection{Representation}
A physical object can be represented in the abstract domain, through a representational relationship, {\fontfamily{cmr-12}\selectfont R}. For example, a physical on-off switch can be represented in the abstract domain by a binary digit, shown in Figure \ref{fig:ar1}. The general representation relation between a physical object, \textbf{p}, and an abstract object, $m_\textbf{p}$, is through a directed map {\fontfamily{cmr-12}\selectfont R} $: \textbf{p} \rightarrow m_{\textbf{p}}$. The abstract object, $m_{\textbf{p}}$ is said to be an abstract \textit{representation} of the physical object \textbf{p}. It it very important to keep in mind that the representation relation is not a mathematical function or a logical relation, but rather a modelling relation that bridges the divide between the physical and the abstract spaces. 




\subsubsection{Physical and Abstract Evolution}
In the abstract domain, there can be an evolution or process, \textit{C : $m_\textbf{p} \rightarrow m_\textbf{p}'$}, that changes an abstract object $m_\textbf{p}$ to another abstract object $m_\textbf{p}'$. Similarly in the physical domain, a corresponding evolution ${\textbf{H} : \textbf{p} \rightarrow \textbf{p$'$}}$, changes the physical state \textbf{p} to physical state \textbf{p$'$}. This physical state, \textbf{p$'$} can then be represented as $m_\textbf{p$'$}$, through the representation relation, {\fontfamily{cmr-12}\selectfont R}. These concepts are shown in Figure \ref{fig:commute} (a). 

The two abstract objects, $m_\textbf{p$'$}$ and $m_\textbf{p}'$, lead us to a key concept in AR theory. If $\left| m_\textbf{p$'$} - m_\textbf{p}' \right| \leq \epsilon$,  for some error $\epsilon$ and norm $\left| \right|$, then we can say that the abstract evolution, \textit{C}, and the corresponding physical evolution, \textbf{H}, \textit{commute}. Under the above condition, the two representation relationships, {\fontfamily{cmr-12}\selectfont R}$ : \textbf{p} \rightarrow m_{\textbf{p}}$ and {\fontfamily{cmr-12}\selectfont R}$ : \textbf{p$'$} \rightarrow m_{\textbf{p$'$}}$, and the pair of abstract and physical evolutions \textit{C : $m_\textbf{p} \rightarrow m_\textbf{p}'$} and ${\textbf{H} : \textbf{p} \rightarrow \textbf{p$'$}}$, can be said to form a \textit{commuting diagram}. When a set of abstract and physical objects form a commuting diagram using the representation, {\fontfamily{cmr-12}\selectfont R}, then $m_{\textbf{p}}$ is a \textit{faithful abstract representation} of physical system \textbf{p} for the evolutions \textit{C($m_{\textbf{p}}$)} and \textbf{H}(\textbf{p}). This means we can be confident that the evolution \textit{C} in the abstract domain corresponds to the evolution \textbf{H} in the physical domain.

The implication of commuting diagrams is that the abstract representation of the final state of a physical object (i.e. $m_\textbf{p}'$/$m_\textbf{p$'$}$) can be found either by following the physical evolution and then representing the output abstractly, or by theoretically evolving the representation of the physical state. As an example, consider a commuting diagram in which physical voltages are represented by binary numbers, shown in Figure \ref{fig:commute} (b). Assume that we want to perform an abstract evolution (binary addition). Then this evolution can be performed either in the abstract domain using logic gates, or through physical manipulation of voltages in the physical domain and representing the result abstractly. 

\subsubsection{Instantiation}
The instantiation relationship, {\fontfamily{cmr-12}\selectfont \~R}, can be thought of as the inverse to the representation relation. Just as a physical object can be represented in the abstract domain by the representation relation, the instantiation relation, {\fontfamily{cmr-12}\selectfont \~R}$: m_{\textbf{p}} \rightarrow \textbf{p}$, instantiates an abstract object in the physical domain.

However, unlike representation, the instantiation relation can only exist under specific conditions, as there are many abstract objects that have no physical instantiation. It is necessary for a commuting diagram to exist for a given representation relation, before any attempts can be made to find the inverse instantiation relation (see \cite{Horsman2014}). Finding an instantiation relation is not straightforward and can be thought of as finding a physical system, that when represented abstractly gives the abstract object that we desire. This often requires trial and error, if such an instantiation is even possible. 

Figure \ref{fig:commute} (c) shows the binary addition example, but this time the voltages instantiate binary numbers. The instantiation relation, {\fontfamily{cmr-12}\selectfont \~R}, can be used to change the physical state to \textbf{p}, so that it instantiates the numbers we wish to add. In the abstract domain, the abstract mapping (i.e. mathematical and logical operations) \textit{C}, performs the addition to arrive at the result. Meanwhile, the physical mapping, \textbf{H}, manipulates the voltages to produce the physical result. Using the representation relation, {\fontfamily{cmr-12}\selectfont R}, the abstract representation of \textbf{p$'$} is found. If we have confidence that the representation is a faithful abstract representation and also that the instantiation relation is correct, then the outcome of the abstract and the physical evolutions should be the same.

\subsubsection{Compute Cycle}
The previous example describes a computer performing a parallel operation in the abstract and physical domains. However, the most interesting use of a computer is when the abstract mapping \textit{C} is unknown and we can use the computer to solve an abstract problem, shown in Figure \ref{fig:commute} (d). Provided that we are confident in the capabilities of the computer, we can use the computer to find the solution. 

The full compute cycle is as follows:

\begin{align*}
& m_\textbf{p} \overset{{\fontfamily{cmr-12}\selectfont \tilde{R}}}{\longrightarrow} \textbf{p} \overset{\textbf{H}}{\longrightarrow} \textbf{p$'$} \overset{{\fontfamily{cmr-12}\selectfont R}}{\longrightarrow}[m_\textbf{p$'$} \approx m_\textbf{p}']
\end{align*}

Thus representation and instantiation enable physical computing resources to implement abstract objects and operations, which can be called abstract resources. 

\begin{figure}[!t]
\centering
\includegraphics[width=0.95\columnwidth]{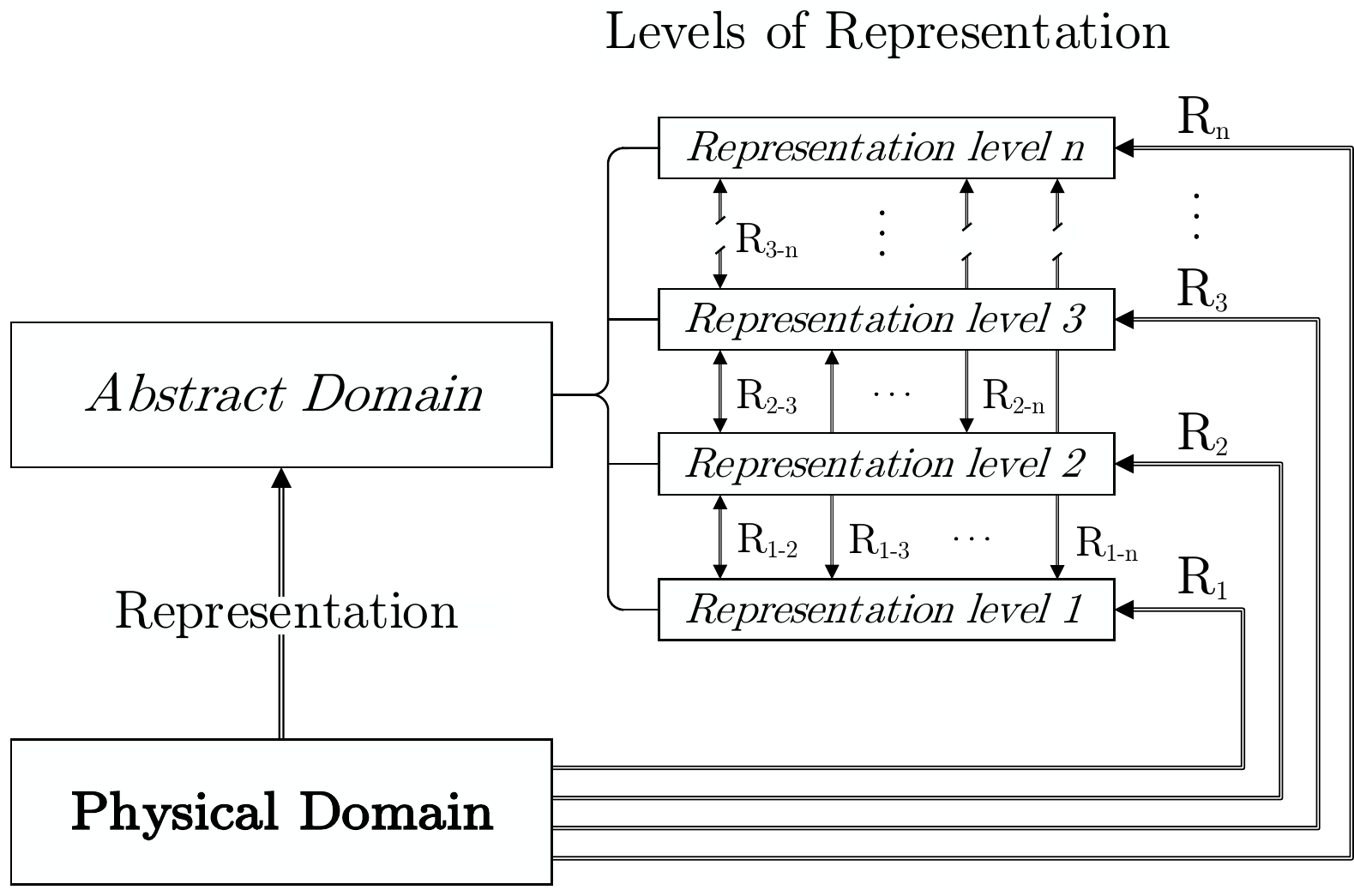}
  \caption{The abstract domain can be divided into many levels of representation, since abstract objects can represent other abstract objects. In this case there are $n$ representation levels, ordered arbitrarily, but there can be an infinite number of levels. Each representation level can represent the physical domain directly, through unidirectional representations {\fontfamily{cmr-12}\selectfont R$_1$}, {\fontfamily{cmr-12}\selectfont R$_2$}, etc., or can represent another representation level, through the bidirectional representations {\fontfamily{cmr-12}\selectfont R$_{1\text{-}2}$}, {\fontfamily{cmr-12}\selectfont R$_{2\text{-}3}$}, etc. These representations are bidirectional to show that any representation level can represent another level. This image shows the theoretical representation relations, which does not mean that these relations will all exist in practice.}
  \label{fig:lor}
\end{figure}

\vspace{-0.5em}
\subsection{Levels of Representation}
\label{subsect:abstraction}


We extend AR theory presented in \cite{Horsman2014} by focusing on the abstract domain and examining abstract objects in more detail. Most importantly, we observe that objects in the abstract domain, which represent physical objects, can in turn be represented by \textit{different} abstract objects. The same is true for instantiation; objects in the abstract domain, which are instantiated in the physical domain, can instantiate \textit{different} abstract objects. For the purpose of brevity, from here on we only discuss representation and imply that the same is true for instantiation.


In essence there are many different representations that can be used. We use the term `levels of representation' to capture this idea and this is depicted in Figure \ref{fig:lor}. In this figure there are $n$ different levels of representation in the abstract domain. We number these representation levels from 1 to $n$, but the numbering is arbitrary and is only an identifier.

It is important to note that the representation used is an arbitrary design choice, and that it is possible to represent the physical domain at any representation level. The unidirectional arrows, {\fontfamily{cmr-12}\selectfont R$_1$}, {\fontfamily{cmr-12}\selectfont R$_2$},  etc., show the one-way representation relation from the physical to the abstract domain as seen previously. 

Similarly, any representation level can represent another level, shown in the figure through the two-way representation relations {\fontfamily{cmr-12}\selectfont R$_{1\text{-}2}$}, {\fontfamily{cmr-12}\selectfont R$_{2\text{-}3}$}, etc.

An important point to observe is that there can be an infinite number of representation levels, as any abstract object can represent another abstract object. However, in practise not every representation level will have a physical representation relation. 




\begin{figure}[!t]
\centering
\subfloat{
	\centering
	\includegraphics[height=0.14\textheight]{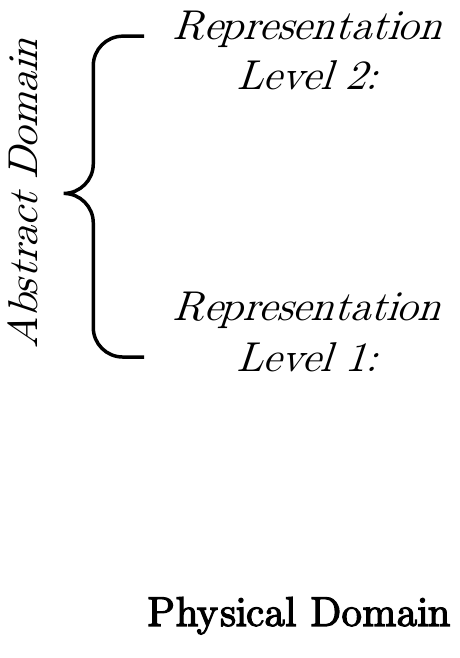}
	\addtocounter{subfigure}{-1}
}
\captionsetup[subfigure]{oneside,margin={0.85cm,0cm}}
\subfloat[]{
	\centering
	\includegraphics[height=0.14\textheight]{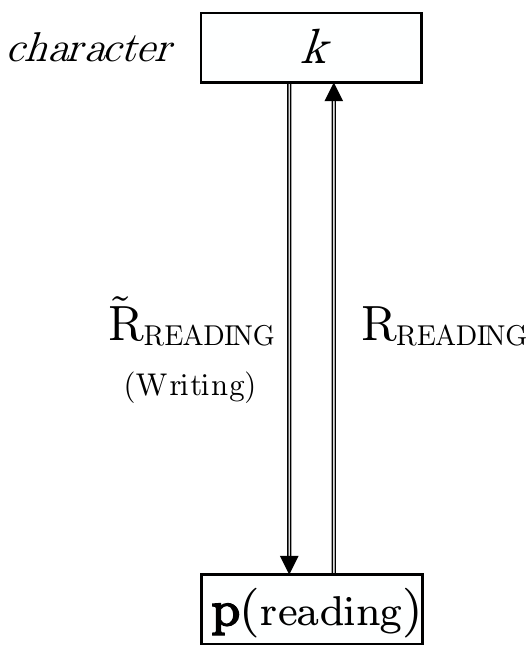}
}
\captionsetup[subfigure]{oneside,margin={0.6cm,0cm}}
\subfloat[]{
	\centering
	\includegraphics[height=0.14\textheight]{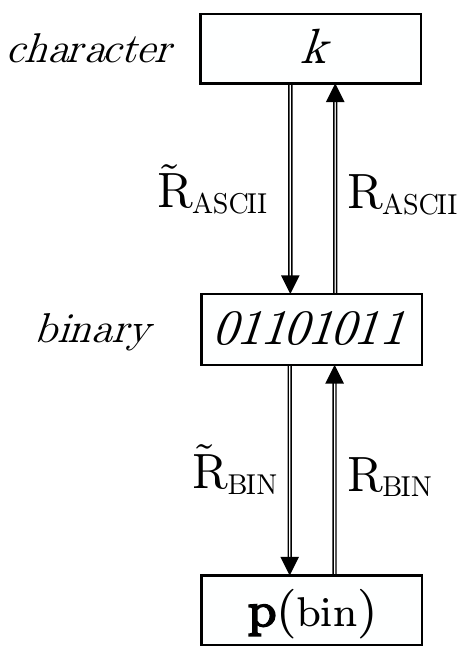}
}
\captionsetup[subfigure]{oneside,margin={0.m,0cm}}
\caption{The number of representation levels is a design choice. For example, the character `k' can (a) directly represent the physical domain, which we commonly know as reading, or (b) represent a binary number, which represents the physical domain. We can speak of one representation level in the first case, the character level, and two representation levels in the second case, the character level and the binary level.}
  \label{fig:levels_respresentation1}
\end{figure}

A practical example is shown in Figure \ref{fig:levels_respresentation1} (a). In this example, a physical object can be represented by the abstract letter `k' directly, and one representation level exists. However, Figure \ref{fig:levels_respresentation1}  (b) has two representation levels. In this case the letter 'k' represents a binary number, which in turn represents the physical domain.

\vspace{-0.5em}

\subsection{Choosing a Hierarchy of Representation Levels}

In theory, there can be an infinite number of representation levels in the abstract domain, with an arbitrary ordering of representation levels. However, in practise it is more useful to computing (and other abstract domain applications) if an ordered hierarchy of representation levels exists. Then it is possible to think of lower representation levels that are more concrete, and higher representation levels that are more abstract. When levels of representation are used in this manner, then we can number the representation levels in order of increasing abstraction. 

One advantage of having an ordered hierarchy of representation levels is that existing physical instantiations for abstract objects can be reused, since it is not an easy task to design a physical instantiation of an abstract system \cite{Horsman2014}. Rather than finding a physical instantiation for an abstract system, an instantiation can be found in terms of an abstract system which already has a physical instantiation \protect\footnote{Which is not an easy task either, but easier than physical instantiation}. 

\begin{figure}[!t]
\centering
\captionsetup[subfigure]{oneside,margin={0.7cm,0cm}}
\subfloat[]{
	\centering
	\includegraphics[height=0.14\textheight]{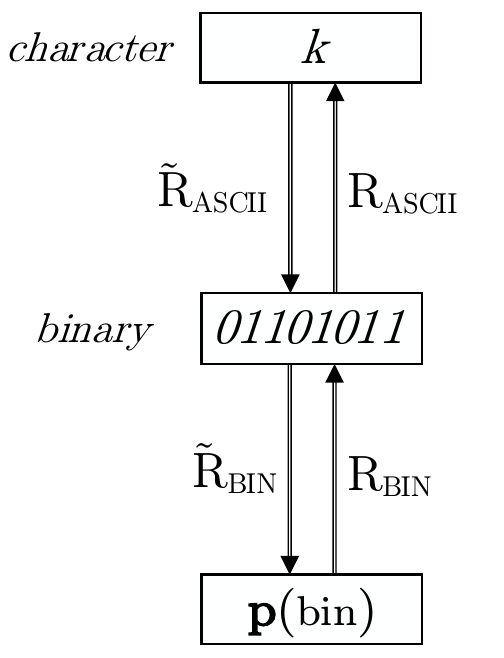}
}
\captionsetup[subfigure]{oneside,margin={0.85cm,0cm}}
\subfloat[]{
	\centering
	\includegraphics[height=0.14\textheight]{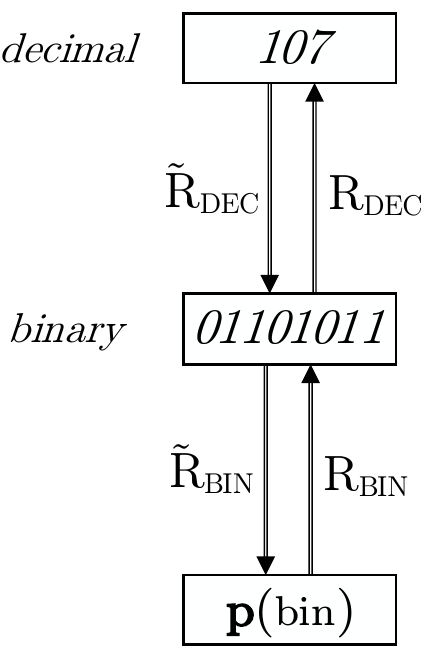}
}
\captionsetup[subfigure]{oneside,margin={0.7cm,0cm}}
\subfloat[]{
	\centering
	\includegraphics[height=0.14\textheight]{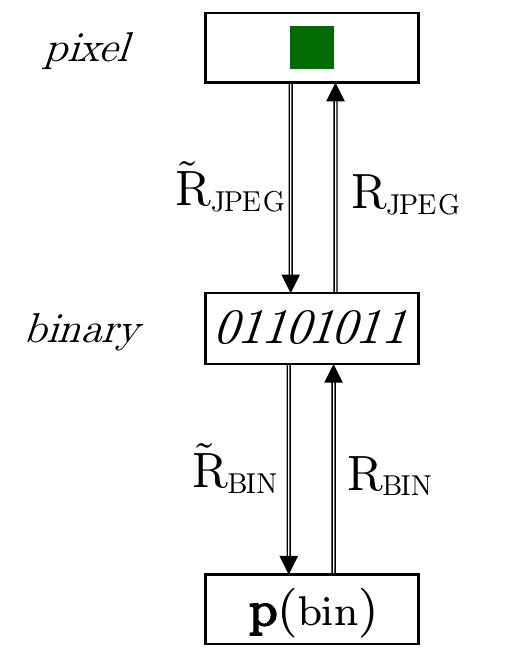}
}
\captionsetup[subfigure]{oneside,margin={0cm,0cm}}
\caption{Multiple levels of representation enable a single physical instantiation to be reused for several different abstract objects. In this example, depending on the representation/instantiation, a binary number could instantiate (a) a character, (b) a decimal number, or (c) a pixel in an image. }
  \label{fig:levels_respresentation2}
\end{figure}

Figure \ref{fig:levels_respresentation2} shows the advantage of multiple representation levels. In this case, the existing physical instantiation of binary numbers can be used to instantiate additional abstract objects. Many types of data such as integers, characters, volume levels, image brightness, and instructions, can be instantiated in computing using bits \cite{Binary}. Thus, using multiple levels of representation provides flexibility and easier instantiation. In this example, the binary representation level can be considered more concrete, and the character/decimal/pixel representation level can be considered more abstract. 

\subsection{Hardware and Software through Representation}

The concept of levels of representation fits in very well with the idea of ``hardware''  and ``software''. We define hardware as a compute cycle in which the representation/instantiation occurs between the physical and the abstract domain. Software, in contrast, is defined as a compute cycle in which the representation/instantiation is completely in the abstract domain. 

We observe that the concept of representing the physical domain at any representation level (for example in Figure 4.) is consistent with the principle of equivalence of hardware and software, which states: 
\begin{quote}
Hardware and software are logically equivalent. Any operation performed by software can also be built directly into the hardware and any instruction executed by the hardware can also be simulated in software. \cite{Tanenbaum}
\end{quote} 

The need for this distinction between hardware and software will be useful later in this paper (for example in section \ref{subsect:nodes}).

\subsection{How Representation applies to Virtualization}

We have examined abstraction and representation, firstly to clarify the distinction between these concepts, and secondly because representation plays a role in virtualization in the following ways:

\begin{enumerate}
\item Virtualization must always be done in the abstract domain. The reason for this is that physical resources cannot be shared or combined without modifying their physical properties in some way. For example, it is not possible to split one processor physically to create multiple virtual processors, however it is possible to split a representation of a processor in the abstract domain. Virtualization allows abstract resources to be reorganised in a manner that is is not limited by the underlying physical resources. 
\item Since there can be many different representation levels in the abstract domain, virtualization can be performed at any of these representation levels and virtualization happens \textit{within} a representation level. However, virtualization does not happen across a representation level - that is the act of virtualization does not change abstract resources from one representation level to another representation level. As we will see later, it is important to know the representation level when virtualizing abstract resources.
\end{enumerate}

\section{Virtualization Theory}
\label{sect:theory}

Having gained a better understanding of abstraction and representation, in this section we identify universal concepts of virtualization, and propose a theory of virtualization. Although some of the concepts have been mentioned before in the literature, to the best of our knowledge this is the first time that these concepts are brought together into a unified theory. 



\subsection{Virtualization as Resource Mapping}
\label{subsect:mapping}

As we saw earlier, resources in the abstract domain are representations of the physical domain. To be consistent with the language that is typically used in virtualization, we refer to these resources as \textit{real resources} (RR). Although the term ``real'' is used, these resources are in the abstract domain, and are not physical resources. Also note that the act of representation is not virtualization, rather representation is a prerequisite for virtualization. 

\begin{figure}[t]
\centering
\includegraphics[width=0.495\textwidth]{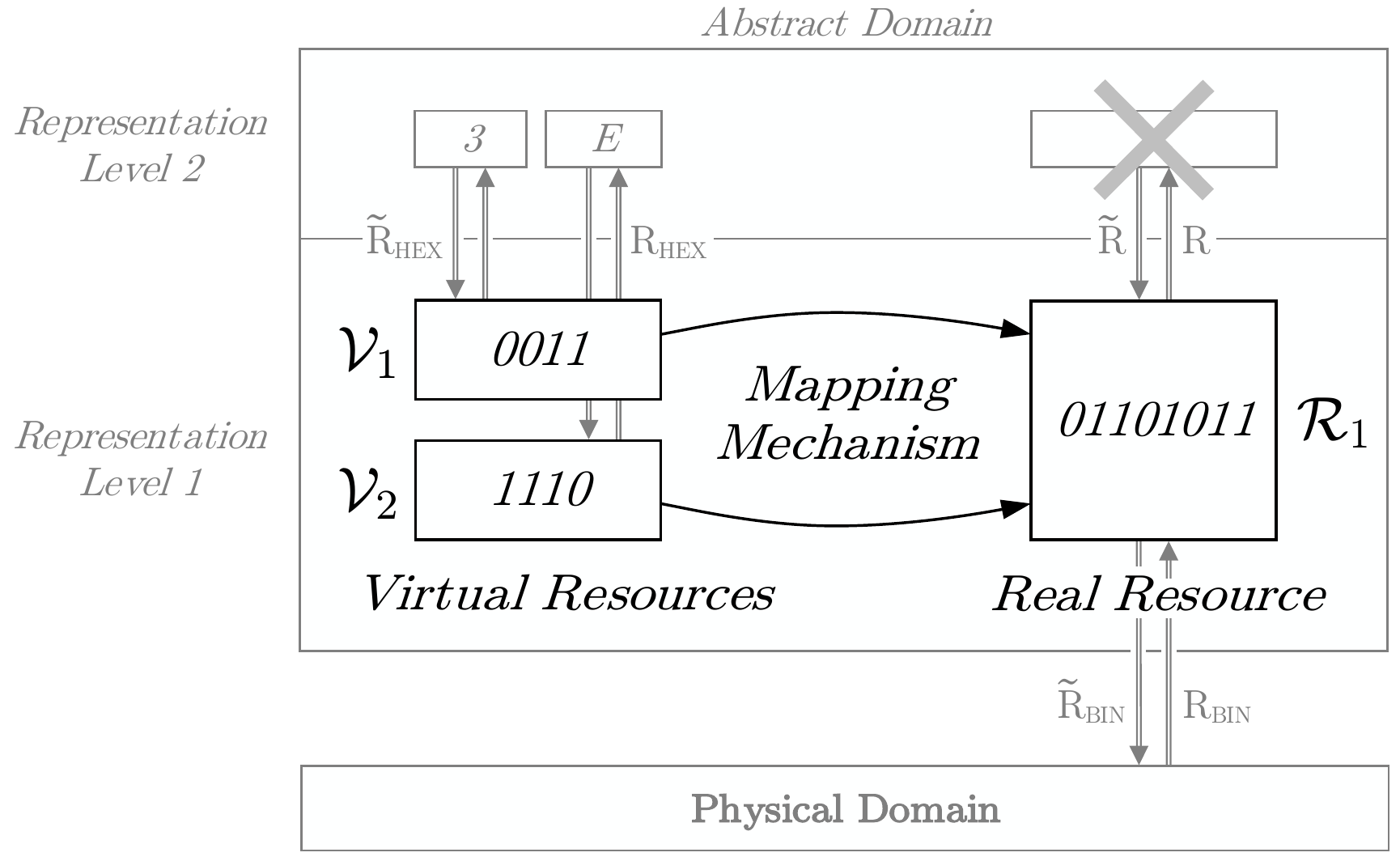}
\caption{Representation allows physical resources to implement abstract resources, while virtualization allows abstract resources to be owned and used at tailor-made quantities. In this example the 8-bit abstract resource $\mathcal{R}_1$ is not an efficient resource to instantiate a hexadecimal number, as hexadecimal numbers only require 4-bits. Mapping two 4-bit virtual resources, $\mathcal{V}_1$, and $\mathcal{V}_2$, to $\mathcal{R}_1$ using a mapping mechanism is a more efficient use of this resource. The virtual resources $\mathcal{V}_1$, and $\mathcal{V}_2$ can then be used to instantiate hexadecimal numbers using the instantiation relation {\fontfamily{cmr-12}\selectfont \~R$_{\textrm{HEX}}$}. The real resource cannot be used to instantiate any higher representation level, shown by the cross in the figure. }
\label{fig:what_is_virtualization}
\end{figure}

Real resources in the abstract domain can subsequently be virtualized. Virtualization is always performed in the abstract domain and at a specific representation level. Virtualization is a resource mapping which can alter the quantity of resources in some dimension(s). The resources after the virtualization process has occurred are referred to as \textit{virtual resources} (VR), as shown in Figure \ref{fig:what_is_virtualization}. Virtual resources appear to be the same type of resources as real resources, but can be altered in quantity in some way. This altering allows abstract resources to be used in a flexible manner, not limited by the underlying representation of the physical domain. However, when virtual resources are mapped to real resources, the real resources cannot be used for any other purpose.

Virtual resources are used as if they were real resources, and it should not be possible to perceive any difference between virtual resources and real resources. In this paper we consider that virtual resources are offered to one or multiple users. The term `users' in this case refers to independent agents, that make decisions on resource use independently. Virtual resources can be offered to different users, and each user has the illusion of full ownership of the resources, meaning that virtual resources can be used for differing purposes. 

In Figure \ref{fig:what_is_virtualization}, the real resource, $\mathcal{R}_1$, is an 8-bit number. This 8-bit number could be used to instantiate a hexadecimal number, or any other instantiation that uses 8 or less bits. However, a hexadecimal number only requires 4 bits to be instantiated, and using $\mathcal{R}_1$ would not be an efficient use of resources. Mapping two 4-bit virtual objects, $\mathcal{V}_1$, and $\mathcal{V}_2$ to $\mathcal{R}_1$, would be more efficient, and allows two hexadecimal (or any other 4-bit instantiation) to be instantiated, using the same resource, $\mathcal{R}_1$. 

Virtualization is always achieved through the use of a \textit{mapping mechanism} (MM) that maps virtual resources to real resources. The mapping mechanism is responsible for presenting the virtual resources as if they were real resources, and for maintaining the isolation between different virtual resources. We call this the \textit{isolation} problem. The mapping mechanism also decides how resources are allocated; in other words deciding how to divide up or combine real resources to create virtual resources. This problem is known as the \textit{embedding} problem. The embedding problem depends greatly on the isolation problem, since the method of isolation determines how the resources can be used. These two problems will be discussed in further detail later.

According to the authors of \cite{Goldberg1973}, and \cite{Denning1970}, the mapping mechanism is simply a function $f$ that maps the set of virtual resources $\mathcal{V}$, to the set of real resources $\mathcal{R}$. The virtual resources can be thought of as the domain of $f$ and the real resources as the codomain of $f$. The mapping function $f$ maps each element in $\mathcal{V}$ to an element in $\mathcal{R}$. 

\vspace{-0.5em}
\begin{align*}
 f: \mathcal{V} \longrightarrow \mathcal{R} \cup \{t\} \\
\end{align*}
\vspace{-2.5em}
\begin{align*}
&  \text{such that if } y  \text{ }\varepsilon  \text{ }\mathcal{V} \text{ and } z  \text{ } \varepsilon \text{ } R \text{ then} \\
&  f(y) =\begin{cases}
&  \hspace*{-3mm} z \text{ if } z \text{ is the real resource for virtual resource } y\\
&  \hspace*{-3mm} t \text{ if } y \text{ does not have a corresponding real resource}
  \end{cases}
\end{align*}

The value $f(y) = t$ causes a trap or fault handling procedure to occur by the mapping mechanism. 

\begin{figure}[!t]
\centering
\includegraphics[width=.99\columnwidth]{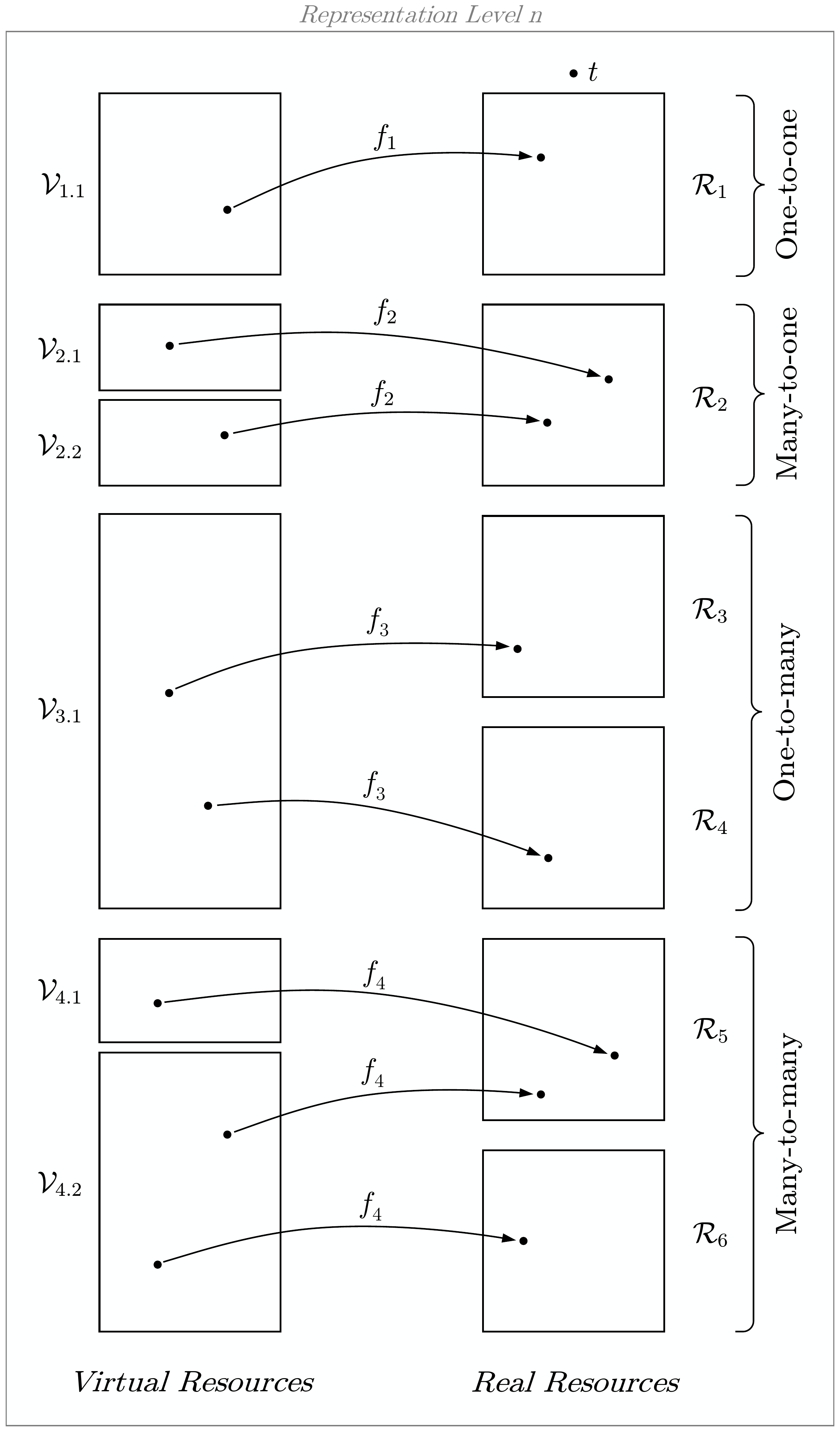}\hfill
  \caption[Types of Virtualization]{In this example, six physical resources are represented as real resources, $\mathcal{R}_{1-6}$, at some representation level $n$. Multiple virtual resources are mapped to the multiple real resources in different ways, showing the four types of virtualization. Many-to-many mapping is the ideal case as it enables abstract resources to be used in the most flexible manner.}
  \label{fig:external}
\end{figure}

In all figures and examples until now, only one physical resource has been considered, represented as one real resource in the abstract domain. However, virtualization can apply to \textit{multiple} real resources, that are representations of \textit{multiple} physical resources. Similarly, there are several ways in which the mapping of virtual to real resources can be done - we identify four types of virtualization. In Figure \ref{fig:external} we show the general types of virtualization, where there can be multiple physical resources, multiple real and virtual resources, and multiple types of mapping. 

The four types of mapping virtual resources, ${\mathcal{V}_n}$, to real resources, ${\mathcal{R}_m}$ are:

\begin{enumerate}
\item One-to-one: Mapping a single virtual resource to a single real resource, to allow for easier management of resources;
\item Many-to-one: Mapping multiple virtual resources to a single real resource such as partitioning a single resource into a number of smaller and more easily accessible resources of same type;
\item One-to-many: Mapping one aggregated virtual resource to several real resources. Used to aggregate many individual components into larger resource pool; and lastly 
\item Many-to-many: Mapping multiple virtual resources to multiple real resources. The combination of aggregating and partitioning resources to create completely customizable resources that can be tailored exactly to requirements.
\end{enumerate} 

Many-to-many can be considered the ideal case, since it enables resources to be used in the most flexible manner.

\begin{figure}
\centering
\vspace{-1ex}
\subfloat[In the first instance of virtualization, many-to-many mapping $f_1$ maps virtual resources $\mathcal{V}_{1.1-1.n}$ to real resources $\mathcal{R}_{1-3}$.]{
	\includegraphics[width=.99\columnwidth]{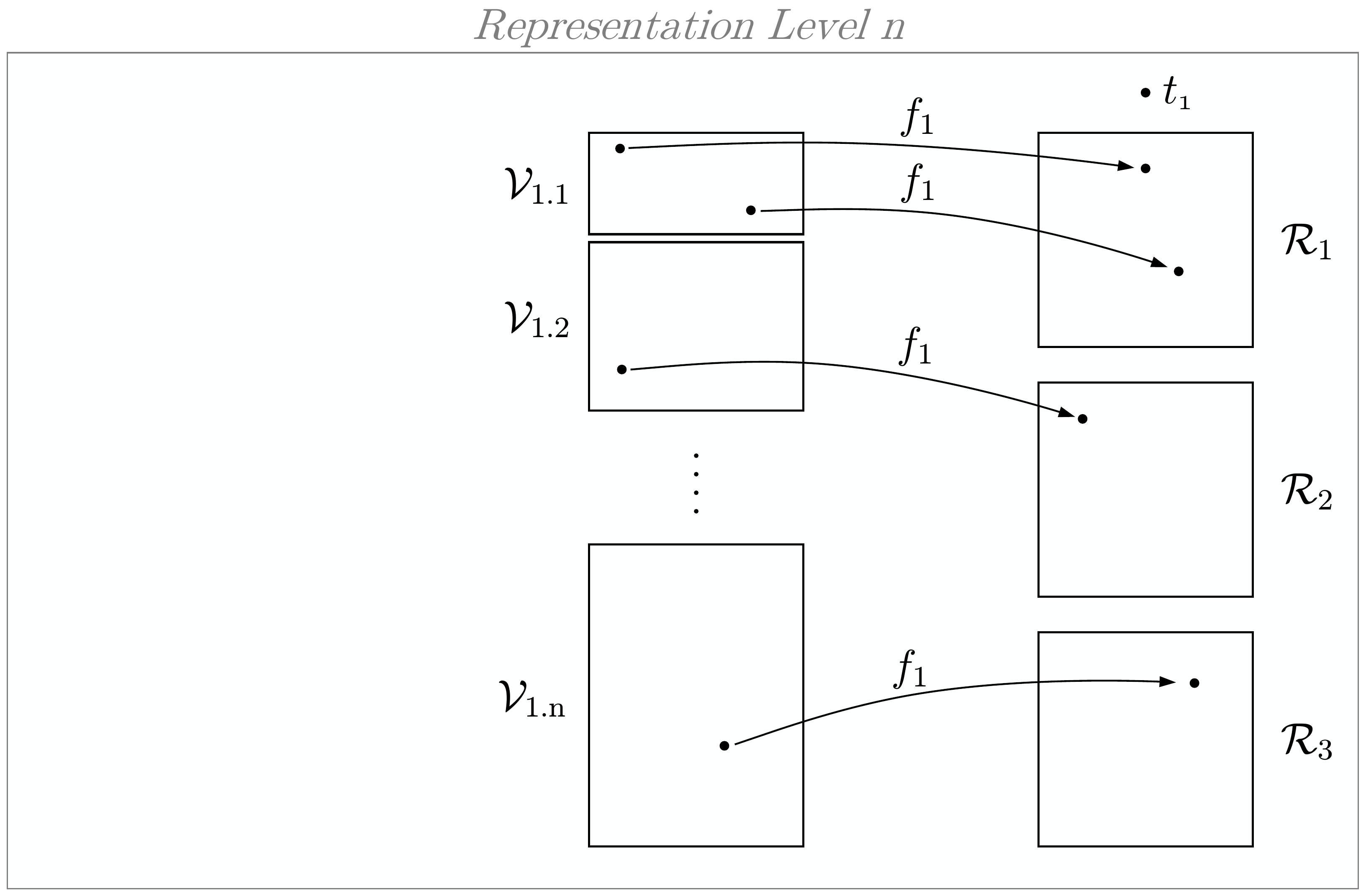}
}

\subfloat[Many-to-one mapping function $f_2$ is an example of recursive virtualization, since it maps virtual resources $\mathcal{V}_{2.1}$ and $\mathcal{V}_{2.2}$ to `real' resource $\mathcal{V}_{1.1}$.]{
	\includegraphics[width=.99\columnwidth]{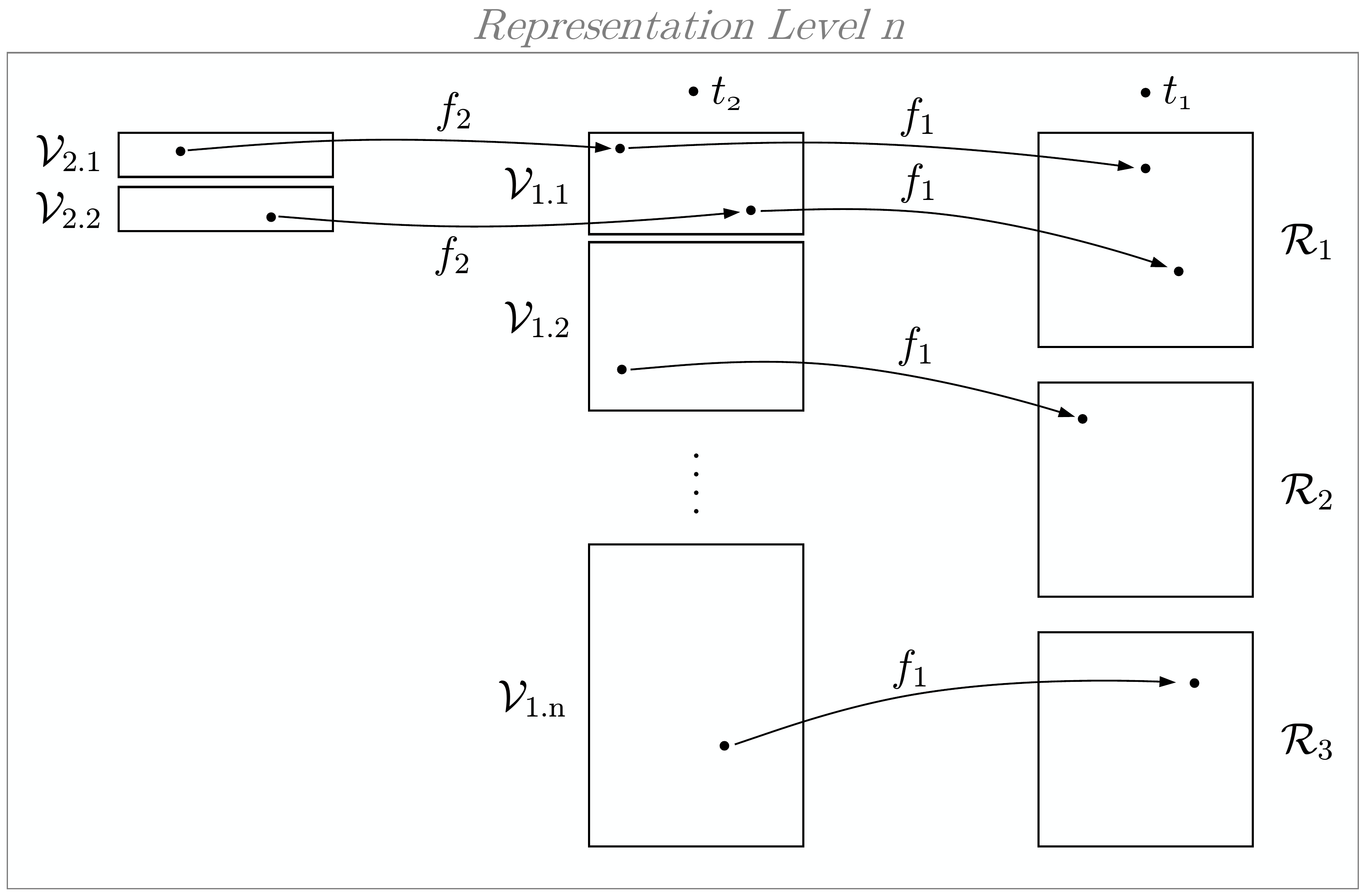}
}


\subfloat[The one-to-many function $f_3$ shows that it could be possible to map virtual resources to a combination of real resources, whether these have been previously virtualized or not. In this case $\mathcal{V}_{3.1}$ is mapped to $\mathcal{R}_4$ and $\mathcal{V}_{1.n}$.]{
	\includegraphics[width=.99\columnwidth]{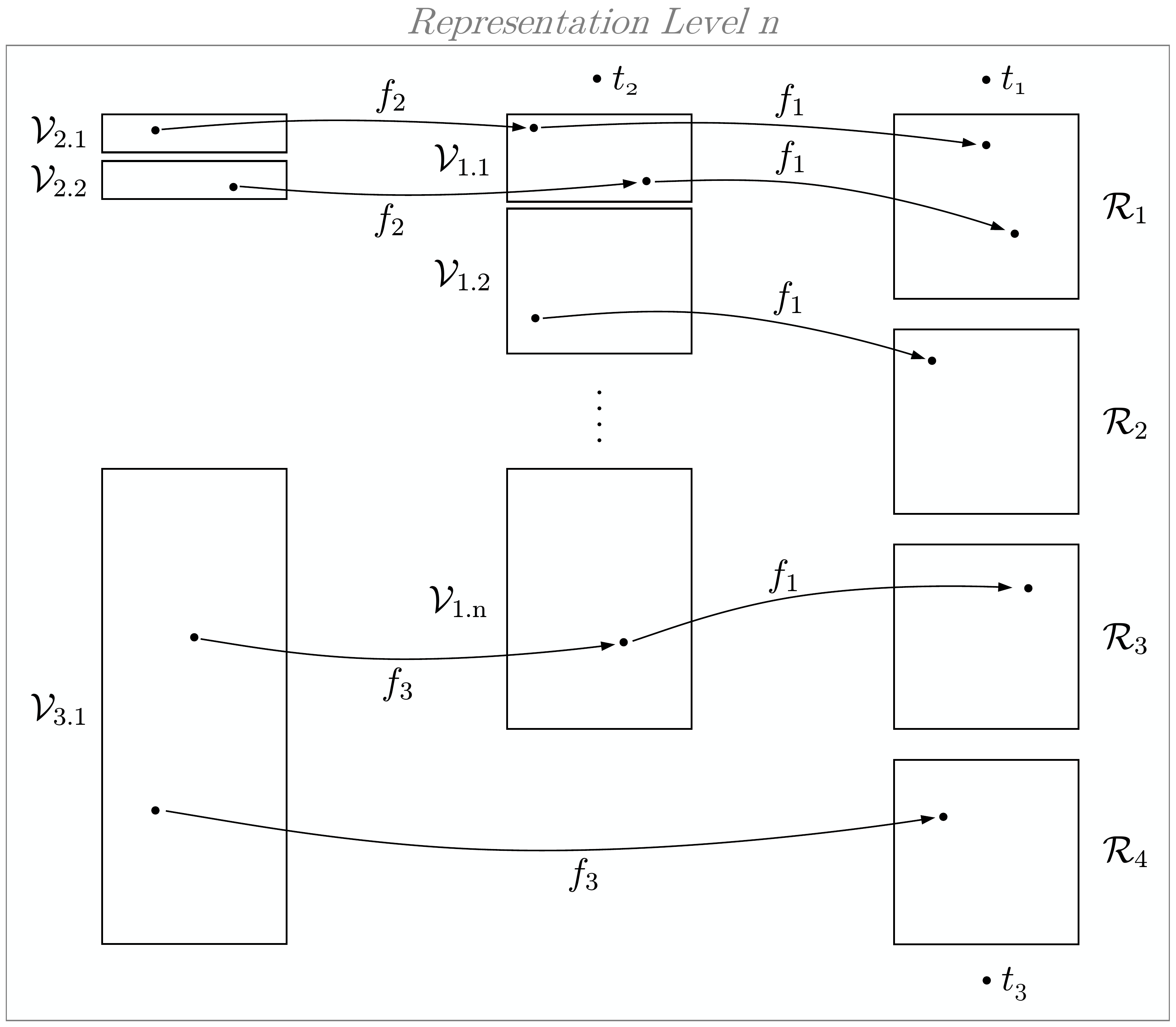}
}	
	
\caption[Recursive Virtualization]{Example of recursive virtualization. $f_1$ is the first instance of virtualization, $f_2$ is the second instance, and $f_3$ is the third. $f_2$ and $f_3$ are recursive virtualization, since they depend on the first instance of virtualization, $f_1$.}
  \label{fig:recursion}
\end{figure}


\subsection{Recursion}
\label{subsect:recursion}

When virtual resources are presented in such a way that they are indistinguishable from real resources, recursion is possible \cite{Popek1974}. Users could choose to virtualize their virtual resources, since they perceive them as real resources. The mapping of resources is now done twice, the first mapping maps the virtual resources received by the user to the real resources, and the second mapping maps virtual resources to virtual resources.

The terms `real' and `virtual' can be confusing when recursion is taken into account. For this reason we always refer to the codomain of the mapping function, i.e. the resources that are being virtualized, as `real' resources, even though these resources might already have been virtualized by a previous virtualization instance. 




The resource mapping function, $f$, described above can be extended directly for recursion by applying the mapping function, $f$, multiple times and interpreting $\mathcal{V}$ and $\mathcal{R}$ as different instances of virtualization. Mapping function $f_1$ maps virtual resources $\mathcal{V}_1$ to real resources $\mathcal{R}$. Now, in a second virtualization instance, $f_2$ maps virtual resources $\mathcal{V}_2$ to `real' resources $\mathcal{V}_1$. 

\begin{align*}
& f_1: \mathcal{V}_1 \longrightarrow \mathcal{R} \hspace{0.6mm} \cup \{t_1\}\\
& f_2: \mathcal{V}_2 \longrightarrow \mathcal{V}_1 \cup \{t_2\}\\
\end{align*}

The real resources for $f_1$ are $\mathcal{R}$, and for $f_2$ the real resources are $\mathcal{V}_1$. Figure \ref{fig:recursion} illustrates several virtualization instances that recursively map virtual resources to real resources. 

We can say that recursion is a requirement for virtualization, since non-recursive virtualization is simply multiplexing \cite{Belpaire1975}. In the case of perfect virtualization, i.e. that virtual resources can be used exactly as real resources and no overhead exists, infinite recursion is possible \cite{Belpaire1975}. 





\subsection{The Isolation Problem}
\label{subsect:isolation}

\begin{figure}[!th]
\centering
\includegraphics[width=.98\columnwidth]{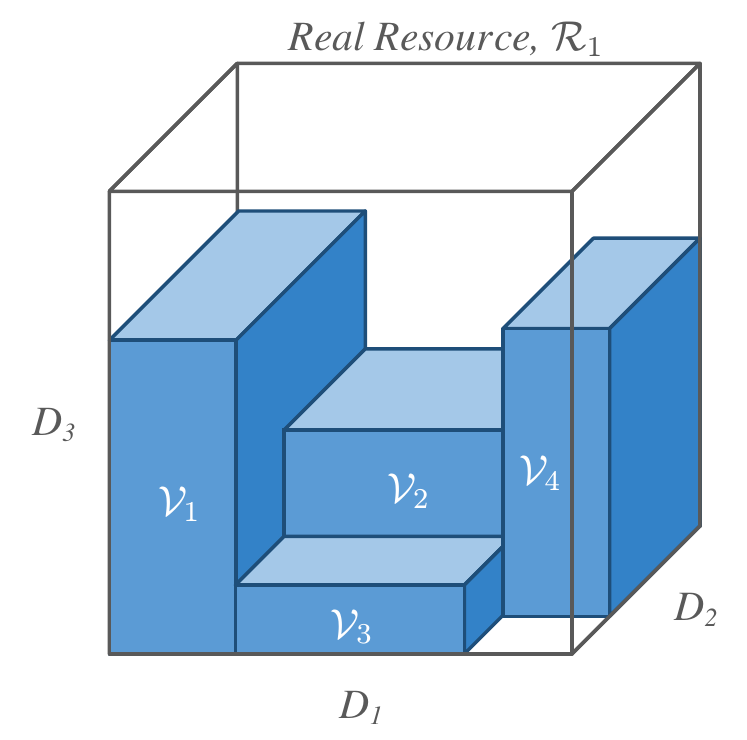}\hfill
  \caption[Virtual Resource Isolation]{The real resource, $R_1$, can be isolated along three dimensions, $D_1$, $D_2$, and $D_3$. The users of virtual resources, $V_1$, $V_2$, $V_3$, and $V_4$, are only aware of and can only access each of their individual resources, which appear to them as real resources.}
  \label{fig:dimensions}
\end{figure}

The isolation problem is the problem of choosing how to create virtual resources, and how to maintain independence between them. It should be impossible for VRs to interact with other VRs in any manner. By isolating along one or multiple dimensions of the real resources, each virtual resource user is only aware of its own virtual resources and can only use those resources, and thus it cannot interfere with other virtual resource users \cite{Park2009} \cite{Sachs2008}. This is illustrated in Figure \ref{fig:dimensions}. The term `dimension' refers to a measurable feature of a resource. 

\begin{figure}[!th]
\centering
\includegraphics[width=.98\columnwidth]{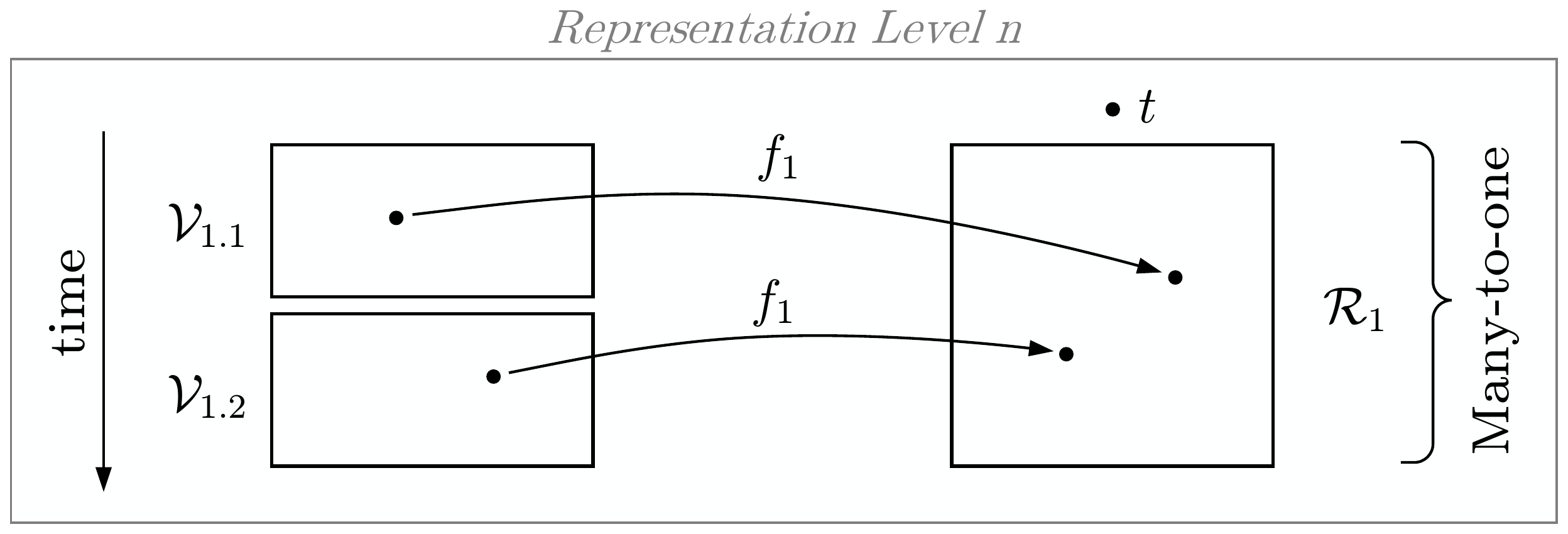}\hfill
  \caption{A many-to-one mapping using one dimension (time) for isolation.}
  \label{fig:iso_map}
\end{figure}

Thus we add the isolation dimension(s) to the mapping function: 

\begin{align*}
 f_1: \mathcal{V}_1 \overset{T}{\longrightarrow} \mathcal{R} \cup \{t_1\}\\
\end{align*}

\noindent for example using the time dimension. An example of how this applies is shown in Figure \ref{fig:iso_map}.


However, the ability to isolate using a particular dimension depends on the technical capability of the mapping mechanism. The granularity used by the isolation process is very important as the user of the virtual resources must not be able to perceive any difference between the virtual resource and the real resource. For example, in processing virtualization, processing resources can be isolated in the time dimension. However, the timescale used by the isolation process is so small (smaller than human reaction time of approximately 0.2 seconds) that the user can use the virtual resources as if they were real resources. If the timescale used by the isolation process was too large, say one hour, then the users of the virtual resources would realise that the resources they are using are not real resources.

Although the isolation problem is a prerequisite to the embedding problem, and influences the embedding problem greatly, in the literature the isolation problem has received significantly less attention compared to the embedding problem.


\subsection{The Embedding Problem}
\label{subsect:embedding}

The problem of deciding how to map virtual resources to real resources is also known as the embedding problem. Essentially this is a resource allocation problem, which is the distribution of scarce resources to competing users. There are several problems that can be considered: 

\begin{enumerate}
\item The first case is that there might not be enough resources to satisfy all of the users' requests; 
\item The second case is that the users can request different quantities of resources; and
\item The third case is that each of the resources can be of unequal value.
\end{enumerate} 

These three problems are not mutually exclusive and often occur simultaneously. Depending on the objective(s) that the resource owner wants to achieve, different metrics can be used to determine the optimal resource allocation. Even when there are enough resources to satisfy all of the users' requests and a potential solution exists, resource allocation can be a complex problem to solve.

\begin{figure}[!t]
\centering
\captionsetup{width=.46\columnwidth, font=small}
\subfloat[The first type of problem occurs when there are more user requests than resources. At least one user will not receive resources.]{
	\centering
	\includegraphics[width=.47\columnwidth]{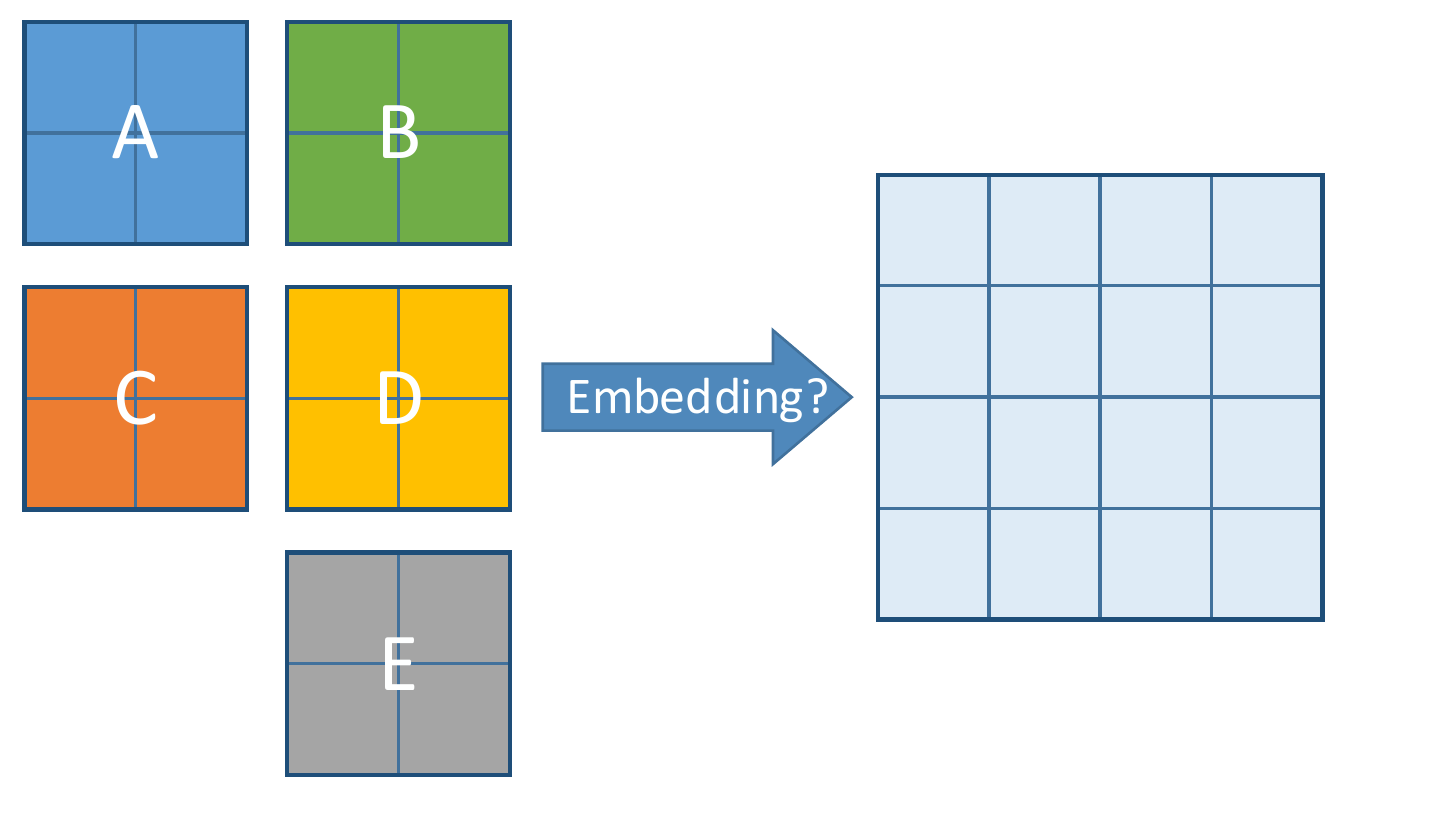}
}
\subfloat[Another type of problem can happen when users have different request sizes.]{
	\centering
	\includegraphics[width=.47\columnwidth]{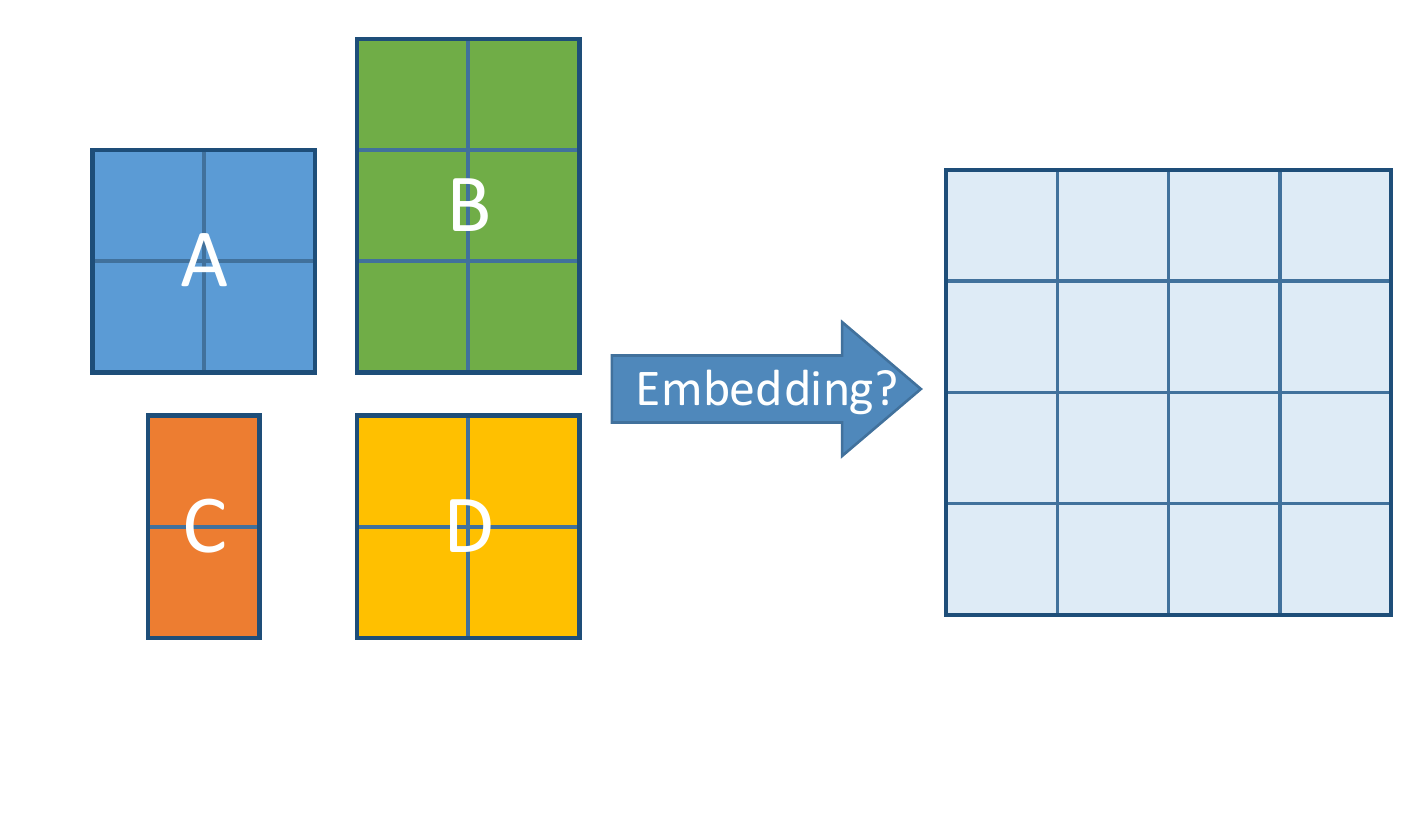}
}
\vskip\baselineskip
\subfloat[A similar problem can occur if the resources are of different sizes.]{
	\centering
	\includegraphics[width=.47\columnwidth]{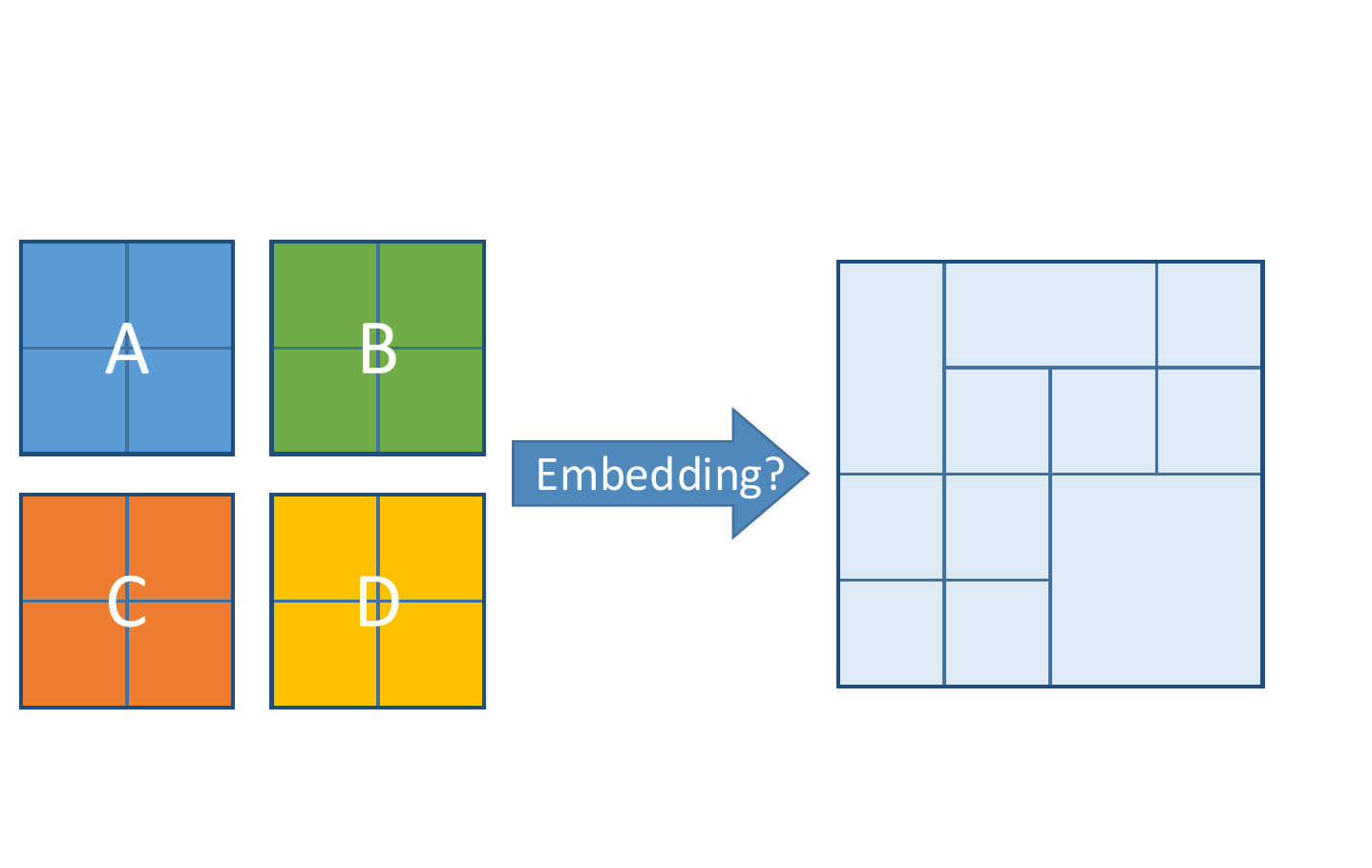}
}
\subfloat[The last example shows that these three problems can occur simultaneously.]{
	\centering
	\includegraphics[width=.47\columnwidth]{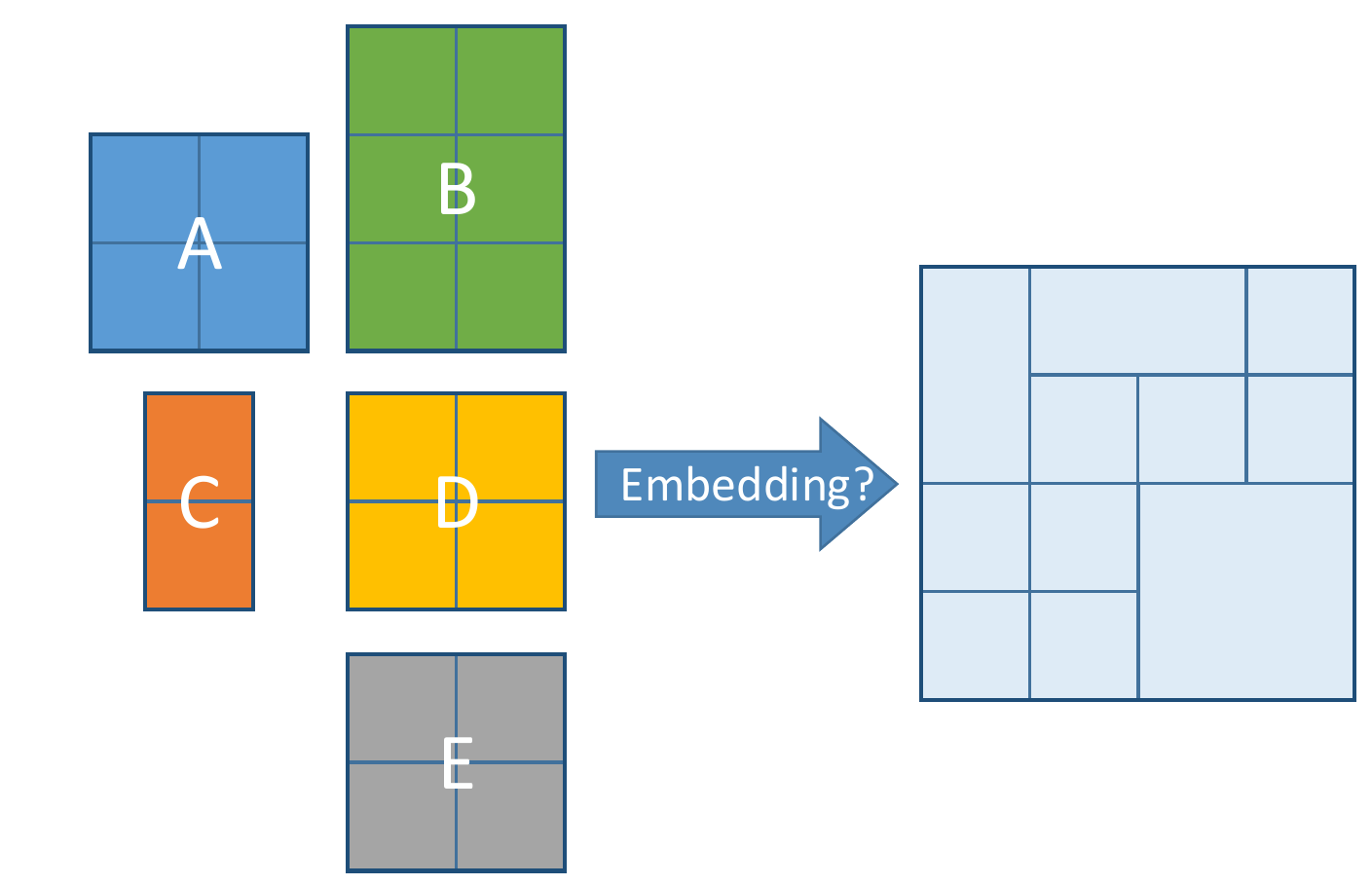}
}
\captionsetup{width=.98\columnwidth}	
\caption[Types of Resource Allocation Problems]{This figure shows resource allocation problems that can happen when users A, B, C, D, and E request two-dimensional resources from a resource owner. There are three different types of problems that can occur, shown in a), b) and c). The three can occur simultaneously as shown in d). Resources can be allocated request by request in sequence, or by considering multiple requests together, which could provide more efficient embedding.}
  \label{fig:embedding}
\end{figure}

In the context of virtualization, users can make requests for virtual resources and an embedding algorithm determines which requests are successful. Figure \ref{fig:embedding} shows the resource allocation problems that can occur when users make requests for two-dimensional sets of resources. 



%


\subsection{Definition of Virtualization}

Now that we have a better understanding of virtualization, we propose the following definition for virtualization: 

\begin{quote}
Virtualization is a resource mapping that occurs in the abstract domain. Virtualization takes places within any one representation level in the abstract domain. The representation level used is a design choice. Abstract resources before virtualization, known as real resources, are limited by the granularity of the underlying physical resources. Abstract resources after virtualization, known as virtual resources, are not, and can be larger or smaller than real resources. Virtual resources are independent and can be allocated simultaneously to multiple users, each with the illusion of full ownership of their resources. Virtualization can always be done recursively, since users perceive no difference between the virtual resources and the real resources.
\end{quote} 

This definition is satisfactory since it addresses the concepts of 
\begin{enumerate*}
\item abstract resources as representations of physical resources.
\item the splitting and combining of resources,
\item ownership and isolation, 
\item allocation, and
\item recursion.
\end{enumerate*}

\subsection{Validity of the Theory}
\label{subsect:validity}

Although we have attempted to validate virtualization theory as much as possible by referring to previous works, some aspects are new and need verification. One method of verifying the theory is to examine virtualization technologies and to see if the theory holds. As an example we examine virtual memory, one of the first virtualization techniques developed.
 
\begin{figure}[!t]
\centering
\includegraphics[width=.98\columnwidth]{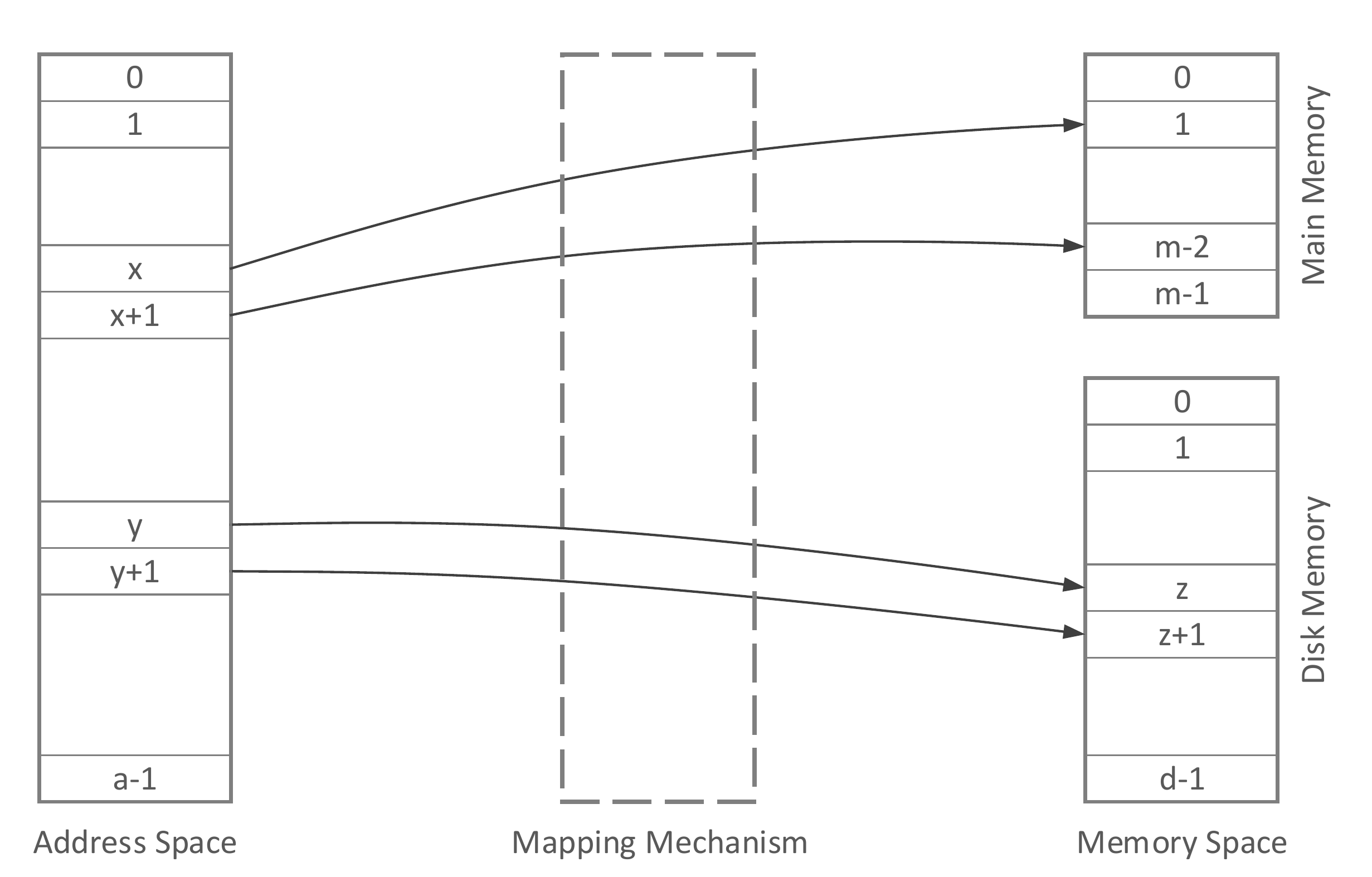}
  \caption{The separation of address and memory spaces allows the address mapping mechanism to combine main memory and disk memory into a one-level storage space. The address mapping mechanism can also give users the perception of having a unique address space for multitasking}
  \label{fig:virtual_memory}
\end{figure}

Virtual memory, or one-level storage as it is also known, was developed in 1961 by the Atlas group to overcome the storage allocation problem of distributing information between main memory and auxiliary memory levels in computers \cite{Kilburn1962}. In a one-level storage system, a distinction is made between the address space, which is the set of identifiers used to refer to information, and the memory space, which is the set of physical memory locations used to store information \cite{Fotheringham1961}. Instead of offering computer programs direct access to the memory space, programs can only access the address space, and a supervisor maps the address space to the memory space. By decoupling the address space in this way from the physical memory, it is possible to \textit{combine} both main memory and auxiliary memory into a single address space, thus offering the illusion of one-level storage, shown in Figure \ref{fig:virtual_memory}.

Let us now examine how virtualization theory applies to virtual memory, summarized in Table \ref{table:example}. Firstly, it is important to identify the abstract resources that are being virtualized. In this case the abstract resources are information storage in the form of bits. Next, it is possible to see that virtual memory is a one-to-many mapping, since it maps one virtual resource (the address space) to many real resources (the main and auxiliary memory spaces) \protect\footnote{With the development of multitasking, in reality virtual memory is now many-to-many, because different programs each have their own address space.}. The isolation dimension is the address space, which is to say the index of locations to store information. In theory recursion is possible in virtual memory, because the resources offered are locations to store information, exactly the same type of resources as non-virtualized memory. In fact, the authors of \cite{Lauer1973} develop such a recursive virtual memory system. As for the embedding problem, some examples of different approaches to embedding are the different paging and segmentation algorithms that have been developed \cite{Denning1970}.


\begin{table}[h]
		\caption{Virtualization Theory Application: Virtual Memory}
	   	\label{table:example}
	    \centering
	   	\begin{tabular}{l l}
	 	\hline \\ [-2ex]
	 	\textbf{Concept} & \textbf{Use in Virtual Memory} \\ 
	 	\hline \\ [-2ex]
	 	\begin{tabular}{@{}l@{}} Abstract resources at \\ a representation level \end{tabular} & \begin{tabular}{@{}l@{}} Location to store bits \end{tabular}	\\ [2ex]
		Mapping  & \begin{tabular}{@{}l@{}} One-to-many / Many-to-many \end{tabular} \\ [0.5ex]
		Isolation & \begin{tabular}{@{}l@{}} Address space - index of storage locations \end{tabular} \\ [0.5ex]
		Recursion  & \begin{tabular}{@{}l@{}} Possible but not very useful in practise, see \cite{Lauer1973} \end{tabular} \\ [0.5ex]
		Embedding & \begin{tabular}{@{}l@{}} Paging algorithms, segmentation algorithms \end{tabular} \\ [0.5ex]
		\hline \\ [-2ex]		
		\end{tabular}
\end{table} 

The example of virtual memory shows how virtualization theory can apply in practise. We see that each aspect of the theory has a practical counterpart in virtual memory. This offers some validation of the theory. In the next section we examine several other virtualization technologies in the context of wireless networks, but for each virtualization technology the same analysis could be performed.

\section{Network Resource Virtualization}
\label{sect:resource_virtualization}

Having developed a theory of virtualization, we now consider how virtualization applies to wireless networks. We define a wireless network as `a set of nodes that can transfer information through links, where some of the links may be wireless in nature'. From this definition, we can deduce that wireless networks consist of two parts: nodes and links. We must first be able to virtualize nodes and links when creating virtual wireless networks, and thus as a first step in the virtualization process, we examine what constitutes nodes and links. We also analyse how node and link functionality can be virtualized.

\subsection{Nodes}
\label{subsect:nodes}

\begin{figure}[!t]
\centering
\subfloat{
	\centering
	\includegraphics[height=0.16\textheight]{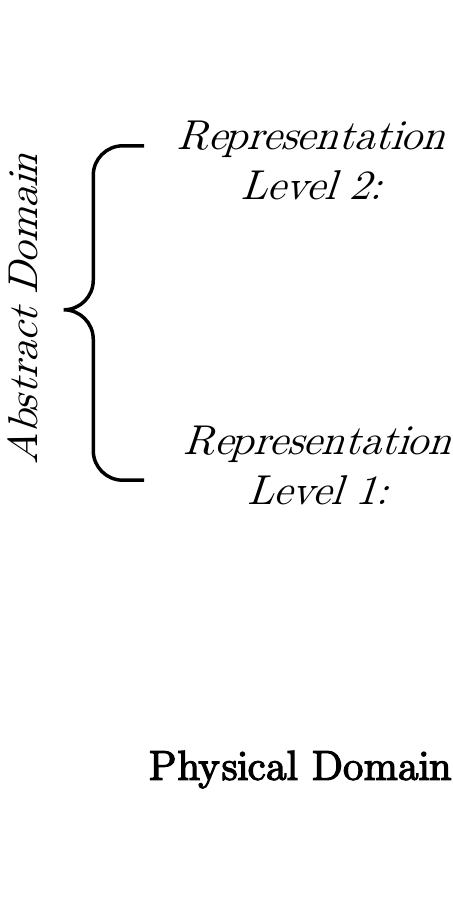}
	\addtocounter{subfigure}{-1}
}
\hfill
\subfloat[ Hardware]{
	\centering
	\includegraphics[height=0.16\textheight]{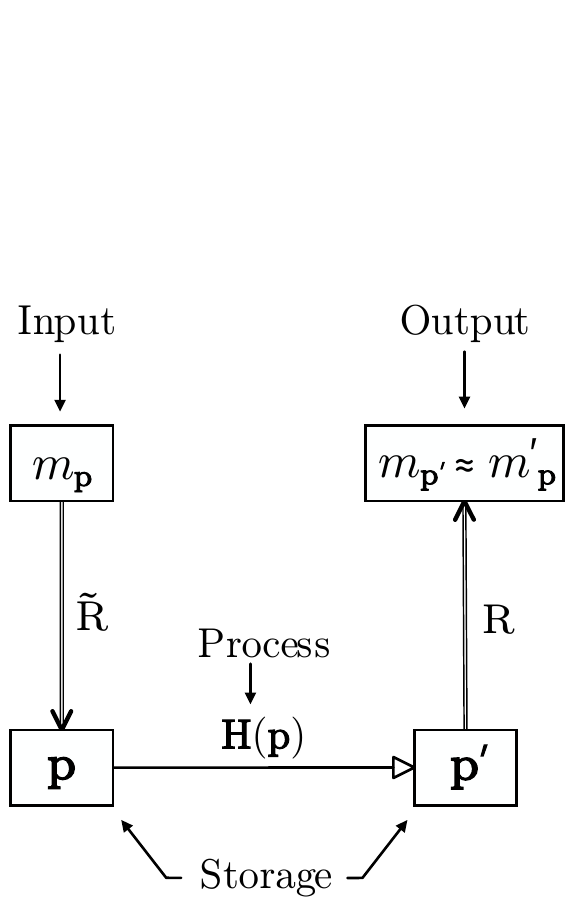}
	
}
\hfill
\subfloat[ Software]{
	\centering
	\includegraphics[height=0.16\textheight]{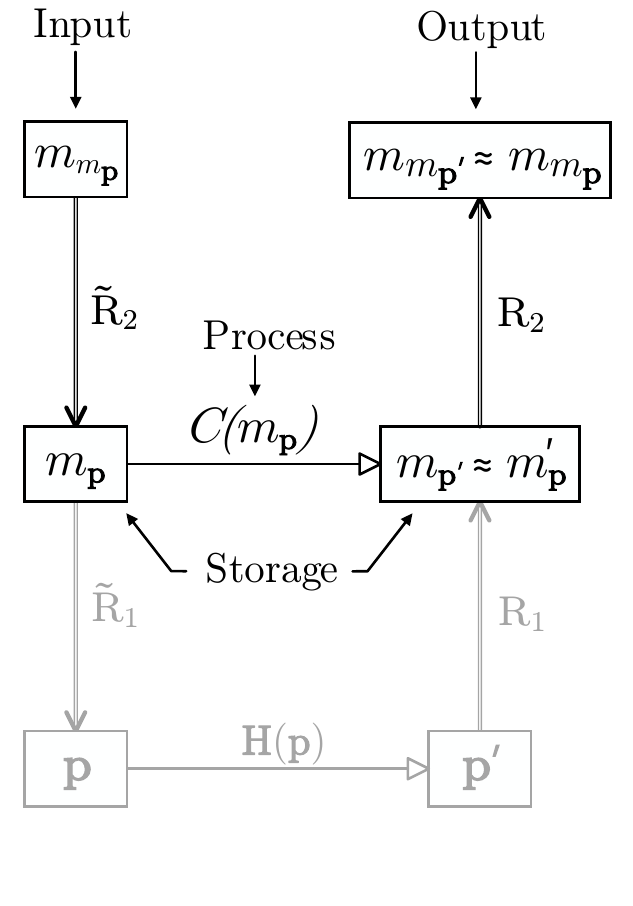}
}
\caption{(a) Model of computer hardware based on AR theory. Computer hardware can be thought of as having four functions, corresponding to the mappings of AR theory and storage of the physical states. (b) Software must be used to enable all four functions to be virtualized, i.e. performing the compute cycle in the abstract domain, as virtualization can only be done in the abstract domain.}
  \label{fig:computer}
\end{figure}

In this paper we consider a node or computer hardware as a physical device that can determine the outcome of abstract operations through physical manipulations \cite{Horsman2014}. Based on the commuting diagram of computer hardware shown in Figure \ref{fig:computer} (a), we can identify four specific functions: \textit{input} (I), i.e. the instantiation of abstract objects in a physical state, \textit{storage} (S) of that physical state in some way, \textit{processing} (P) of the physical state in a way that is commutative to some form of abstract operations, and \textit{output} (O) of result of the physical process in the form of an abstract representation. All functions are required within a node, since, for example, it would be pointless to have processing available but no input, as there would be no information to process. These four functions are consistent with the IPO+S model of computing \cite{Checkland1981}.


Each of these resource types can be represented in the abstract domain and virtualized, which allows resources to be used more efficiently, and also can provide new functionality, such as machine (node) virtualization. However, as mentioned previously, virtualization can only be done in the abstract domain. This means that if all four functions are to be virtualized simultaneously, then it is necessary to introduce an additional level of representation, that is to perform the virtualization in software. We show the show the general compute cycle in the abstract domain (software) in Figure \ref{fig:computer} (b). 

We refer to resources according to their function type, for example we refer to storage resources rather than resources that instantiate storage functionality. Although technically incorrect, this simplification makes it easier to follow. The types of node resource virtualization are discussed briefly below.

 
\subsubsection{Process Virtualization}
\label{subsect:time_sharing}

The idea of process virtualization can be traced back to the concept of compatible time-sharing, first developed at M.I.T. in the early 1960's \cite{Corbatcf1962}. Time-sharing was developed to overcome the limited man-machine interaction of batch processing, which had led to an increase in programming errors and debugging time, as larger and more complex programs were being set \cite{Creasy1981}.

\begin{figure}[!t]
\centering
\includegraphics[width=.67\columnwidth]{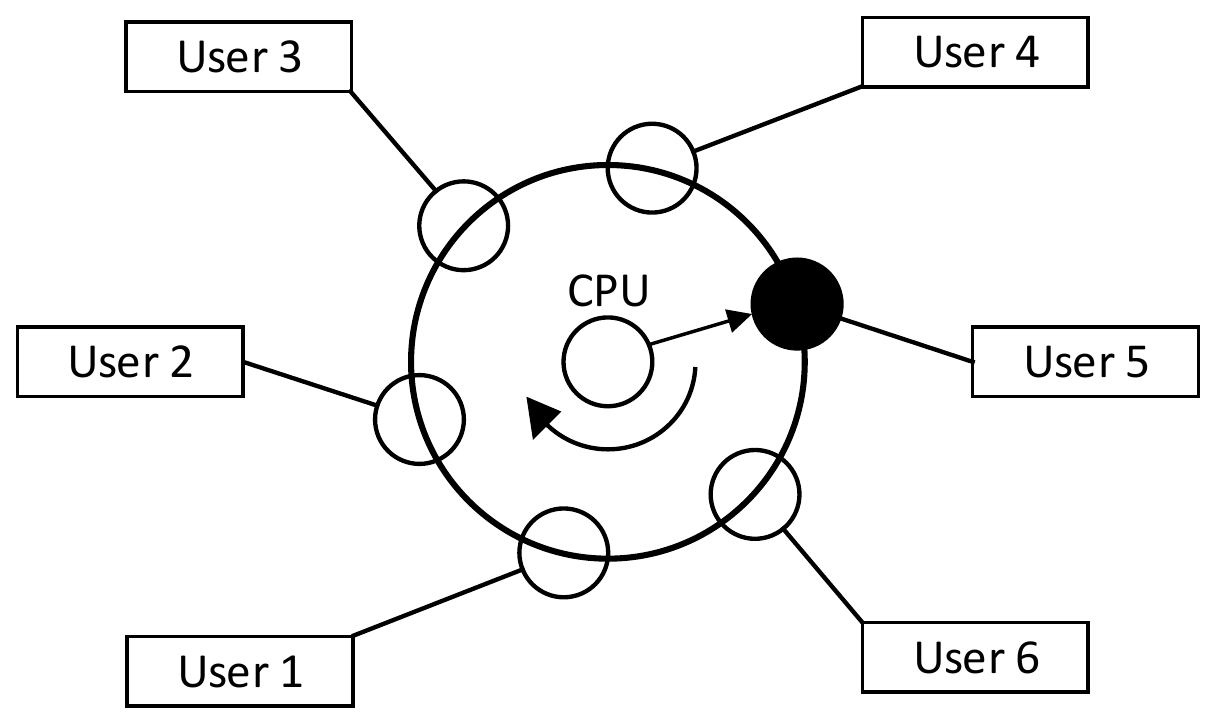}
  \caption{Time-sharing gives several users the perception of having exclusive use of a processor, by giving each user a short burst of computing time. In this example, the CPU is processing user five's burst of time, before it moves on to the next user.}
  \label{fig:time_sharing}
\end{figure}

Time-sharing enables several people to make use of a computer at the same time, shown in Figure \ref{fig:time_sharing} \cite{TimeShare}. Rather than offering users direct access to computing resources, which can lead to serious crashes and memory problems, a supervisor buffers user input, and sequentially runs user programs for small bursts of time. The full sequence of user programs occurs frequently enough (ideally in less than $\sim0.2$ seconds) that a computer appears to be fully responsive to all users. 

By mapping user processing time to burst of machine processor time in this way, and by maintaining strict isolation, users have the perception of exclusive use of dedicated processors. Thus the illusion of multiple `virtual' processors is created. Thus virtualization both increases resource efficiency, and also offers users new or improved functionality. The surveys \cite{McKinney1969, Kleinrock1970, Yashkov2007} provide further information about processing virtualization techniques, and a very interesting and informative documentary on the compatible time-sharing system can be found at \cite{CTSSvideo}.

\subsubsection{Storage Virtualization} 
\label{subsect:storage_virtualization}

One method of storage virtualization is virtual memory, which we discussed in section \ref{subsect:validity}. The use of virtual memory not only automates the storage allocation problem efficiently \cite{Sayre1969}, but also enables machine independence, program modularity, convenient memory addressing, and the capability of handling structured data \cite{Fotheringham1961}, \cite{Denning1970}. For a more detailed view on storage virtualization techniques, see the surveys \cite{Raina1992}, and \cite{Protic1998}.

\subsubsection{Machine (Node) Virtualization}

The development of process and storage virtualization led to new functionality, and computing became thought of as a large system of components serving a community of users, where each of the users could run different programs with different processing, memory, and I/O interaction requirements \cite{Corbato1965}. In such a system, software is commonly split into two classes to avoid system integrity issues: a privileged supervisor (or Operating System) which is presumed to be correct, and a second non-privileged class which is denied any functionality that can cause interference between processes \cite{Critcklow1963}. 

However, this arrangement only allows one privileged supervisor to be run at a time, and incompatible non-privileged programs cannot be run easily \cite{Buzen1973}. Machine virtualization overcomes this problem by constructing simulated copies of the machine, known as virtual machines, and each virtual machine can run a different privileged supervisor \cite{Goldberg1973_2}. A virtual machine monitor (VMM), also known as a hypervisor, isolates process and memory operations for each virtual machine, and maps them to the host machine using time-sharing and virtual memory techniques \cite{Goldberg1974}. Until recently I/O operations had to be trapped and executed by the VMM. 




Advances in machine virtualization, especially in server virtualization have reduced the cost of servers hugely and has led to the widespread adoption of moving computing tasks to the `cloud' \cite{VMware}. The recent development of I/O virtualization (see next part) has allowed full computer node virtualization, consisting of storage, process and I/O virtualization. There are many contexts in which the use of node virtualization is growing such as desktop, application, and user virtualization. More information can be found in the works \cite{Goldberg1974}, \cite{Rose2004}, \cite{Bari2013}, \cite{Chiueh2005}, and \cite{Blenk2016}.

\subsubsection{Input and Output Virtualization}
\label{subsect:io}

One of the problems encountered by early virtual machines was the mapping of input and output (I/O) paths from virtual device addresses to real device addresses, since absolute addressing is required for I/O paths \cite{Buzen1973}. Early Virtual Machine Monitors (VMMs) trapped I/O instructions used by the virtual machines, copied instructions, and `absolutized' them by mapping the virtual addresses to the correct real I/O addresses.
While this solution of emulating I/O devices in software enabled virtual machines to use I/O operations, it was a work-around and not I/O virtualization.


The generalized I/O memory management unit (IOMMU), developed by Intel, is a hardware device that maps virtual device addresses to real ones across isolated partitions, supervised by system software \cite{Abramson2006}. The VMM controls these partitions and thus full virtualization of I/O operations is possible, since virtual I/O operations are mapped directly to the devices. One example of an I/O peripheral that has been virtualized is the Network Interface Card. A survey of I/O virtualization techniques is given in \cite{Zhang2010}.


\subsection{Links}

The function of links is to transfer information between nodes in a reliable manner. Similarly to nodes, links can also be thought of as consisting of abstract objects (i.e. information) and physical resources which instantiate that information and physically send it between nodes. Again, similar to nodes, links can only be virtualized in the abstract domain.

Links can be wired or wireless. However, there are significant differences between wired and wireless links, due to the nature of the physical resources used. Both wired and wireless links represent information using the electromagnetic spectrum, but in the case of wired links, the electromagnetic spectrum is isolated from other links through the use of physical cables. In wireless networks, since all links are broadcast, additional measures must be taken to provide isolation and reliability. 

\subsubsection{Wired Link Virtualization}
\label{subsect:VPNs}

Time-shared systems allowed users to work from remote terminals connected to a mainframe computer, with the perception of working at a personal computer. Connections were implemented either using dial-up lines over the Public Switched Telephone Network (PSTN), or through private lines leased from the PSTN operators. The dial-up lines were much cheaper to use, but suffered from lack of security and functionality, while private lines offered good security and functionality but were expensive. 

Virtual Private Network (VPN) services offer the security and functionality of private lines at much cheaper costs by exploiting the fact that typically communication between nodes in a network only occurs for a small percentage of time. Thus physical links can be time-shared to provide the illusion of private links, known as virtual circuits \cite{Rybczynski1980}. VPNs can be scaled much easier than physical private links, and can also be tailored to suit users' preferences \cite{Metz2003}.

However, although VPNs can isolate different logical networks over a shared infrastructure, they are prone to some limitations. Among these are that the coexistence of different networking solutions is not possible, and also that virtual networks are not fully independent \cite{Carapinha2009}. Another limitation is that broadcasting is not supported in the same way as on native networks. A final limitation is that additional security measures are needed for VPNs which can add overheads, and these and other virtualization overheads can decrease the network reliability, throughput and latency. 

\subsubsection{Wireless Link Virtualization}
Although wired link virtualization has existed for many decades, wireless link virtualization (WLV) is an active research area. As mentioned before, wireless link virtualization is the main focus of this paper. Therefore we briefly introduce wireless link virtualization here, whereas a detailed survey of WLV follows in Sections \ref{sect:survey} and \ref{sect:questions}.

Wireless link virtualization is the process of virtualizing wireless links, creating virtual resources that are isolated and which can use differing technologies and/or configurations independently. Although appearing to be similar to wireless resource sharing, wireless resource sharing is fundamentally different from virtualization, since wireless resource sharing does not create independent resources. In addition, virtualization allows the combination of resources to occur, which resource sharing does not.

There are several complications which exist in wireless link virtualization which do not exist in wired link virtualization due to the difference between the wired medium and the wireless medium. The first difference is that because of the inherent variation of the wireless channel over time, it is not possible to predict the information throughput of wireless links in advance. A second difference is that wireless links are broadcast, and thus have the potential to interfere with any other wireless link, whereas in wired links this does not occur \cite{Park2009}. Another complication for wireless links is that wireless nodes tend to be highly mobile, which means it is harder to predict and provision for the information transfer between nodes.

Because of the above reasons, WLV is a difficult problem and although there has been a significant amount of work on WLV in recent years, there are still many unresolved issues.

\section{Network Virtualization and Wireless Network Virtualization}
\label{sect:nv}

Having examined the virtualization of the resources that make up networks, and also armed with a better understanding of virtualization, we can now examine network and wireless network virtualization. 

\subsection{Network Virtualization}

The idea of Network Virtualization (NV) has arisen as a solution to several problems with the Internet today. As a result of the Internet's success and ubiquity in many areas of life, it has become subject to ossification, since the competing interests of existing stakeholders and the large capital investment needed have led to high resistance to disruptive technologies \cite{Peterson2003}, \cite{Turner2005}. The need for global agreement between competing providers has limited innovation to simple incremental updates, which do not satisfy demand for new services and functionality, or to ad-hoc workarounds, which do satisfy legitimate needs, but violate core Internet design principles, and as such have impaired flexibility, security, reliability, and manageability \cite{Anderson2005}, \cite{Carapinha2009}.

Network virtualization has evolved from the concept of using overlay networks to address these ossification issues. Initially, it was envisaged that overlay networks would be highly programmable platforms for innovation, while simultaneously allowing the existing network infrastructure to be maintained \cite{Andersen2001}, \cite{Peterson2003}. However, later it was realized that overlay networks could not provide the innovation and flexibility desired because of the limitations of the underlying infrastructure, and that it would be necessary to virtualize the underlying network infrastructure \cite{Anderson2005}. 

\begin{figure}[!t]
\centering
\includegraphics[width=.98\columnwidth]{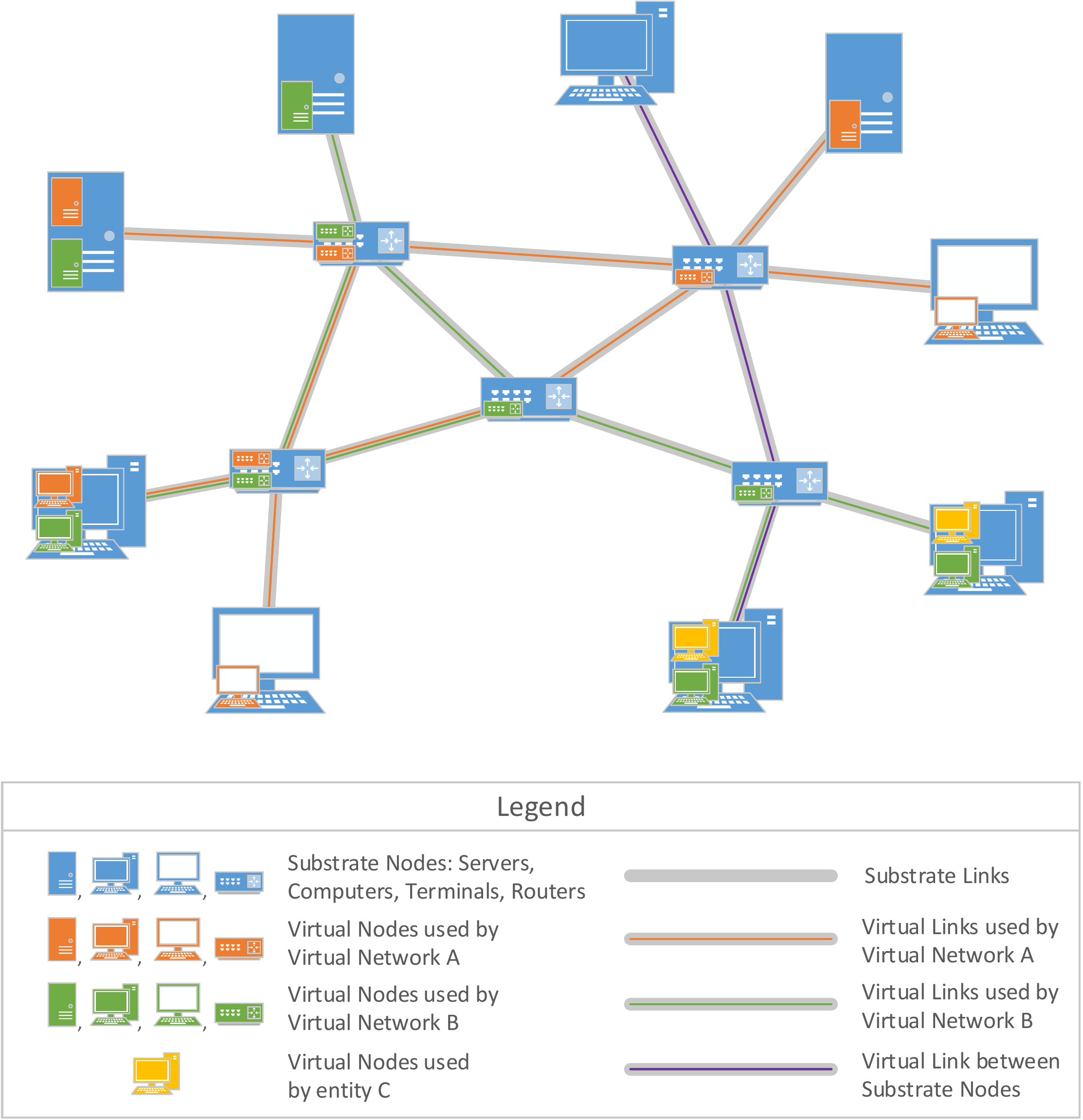}\hfill
  \caption[Node, Link and Network Virtualization]{Network Virtualization: Virtual networks A (orange) and B (green) can coexist simultaneously on the underlying substrate. These networks can offer customized services to their users in a secure and isolated manner. Another entity C requires several computing nodes, but no network connectivity and so node virtualization provides entity C with virtual nodes (yellow). Meanwhile, virtual links or VPNs (purple), can be used to connect several substrate nodes. }
  \label{fig:nv}
\end{figure}

Network virtualization can be considered as using both node and link virtualization to create complete virtual networks (sometimes called Meta-Networks) \cite{Touch2003}, \cite{Turner2005}. Figure \ref{fig:nv} shows the distinction between virtual nodes, virtual links and virtual networks. In addition to offering a potential solution to the Internet impasse, network virtualization also enables the separation of existing Internet providers into the roles of Infrastructure Provider (IP) and Service Provider (SP) \cite{Feamster2007}. As argued in \cite{Turner2005} and \cite{Feamster2007}, there are several reasons why this is beneficial: 

\begin{enumerate}
\item Existing network providers have very few opportunities to distinguish themselves from their competitors and thus they equally have little incentive to develop and deploy new solutions. Decoupling IPs and SPs enables diversity and innovation, leading to better infrastructure and new and improved services for end users.
\item Separating the role of infrastructure and service providers lowers the barriers of entry significantly, since SPs do not need to invest in their own equipment. This leads to increased competition and innovation.
\item A separation of the two roles enables the sharing and aggregation of network infrastructure, leading to increased efficiency and cost savings.
\item In the current network model, new software and network protocols cannot easily be tested, and often not under real traffic conditions. By splitting the roles of SP and IP, new experimental software and protocols can be tested in isolated networks, without affecting existing services. The isolation of networks also enables SPs to customise their networks and achieve better security.
\end{enumerate}

Because of the many potential benefits that can be obtained, network virtualization is an area of ongoing research, and several NV testbeds have been constructed. Initially, testbeds have focused on node virtualization such as PlanetLab \cite{Peterson2003} and GENI \cite{GENI}, but more recently, testbeds such as CABO \cite{Feamster2007} and 4WARD \cite{4WARD} are interested in full network virtualization, i.e. both node and link virtualization combined. There is also significant ongoing research on network virtualization, summarized in the surveys \cite{Chowdhury2010} and \cite{Fischer2013}. 

Two interesting ongoing research topics which are related to network virtualization are Software Defined Networking (SDN) and Network Function Virtualization (NFV). Although these topics are outside the scope of this paper, they should be discussed briefly.

Software defined networking can be seen as ``the separation of [network] forwarding hardware from the control logic'' \cite{Bruno2015}. Thus SDN can be viewed as an enabler for network virtualization, since it allows functionality to be implemented in the abstract, rather than the physical domain. Referring back to Figure \ref{fig:computer}, we can think of SDN as moving functionality, i.e. the compute cycle, from (a) hardware to (b) software. However, it is important to note that the use of SDN does not imply virtualization. There are many benefits to be gained from using SDN in of itself \cite{Kreutz2015}.

The idea behind network function virtualization is to decompose a given network service into a set of functions which can be implemented on commodity hardware through software virtualization techniques \cite{Mijumbi2015}, \cite{Han2015}. NFV can be considered as a form of node virtualization, albeit specifically tailored to networking. For example, parts of the mobile core network such as the mobility management entity, the home subscriber server, and many other functions could be virtualized to enable flexible and dynamic operation. NFV represents a significant step towards network virtualization, as techniques from NFV could be used in network virtualization.




\subsection{Wireless Network Virtualization}


Wireless network virtualization has been proposed as an extension of network virtualization to wireless networks, providing similar potential benefits in terms of flexibility and efficiency \cite{Chandra2004}, \cite{Yang2014}. Specifically, it is hoped that WNV can increase resource efficiency and flexibility for the problems of 1) the continuing rapid growth of demand for wireless services, 2) the ever-greater demand for diversity of services, and 3) the increasing costs of wireless infrastructure. 

\begin{figure}[!t]
\centering
\includegraphics[width=.98\columnwidth]{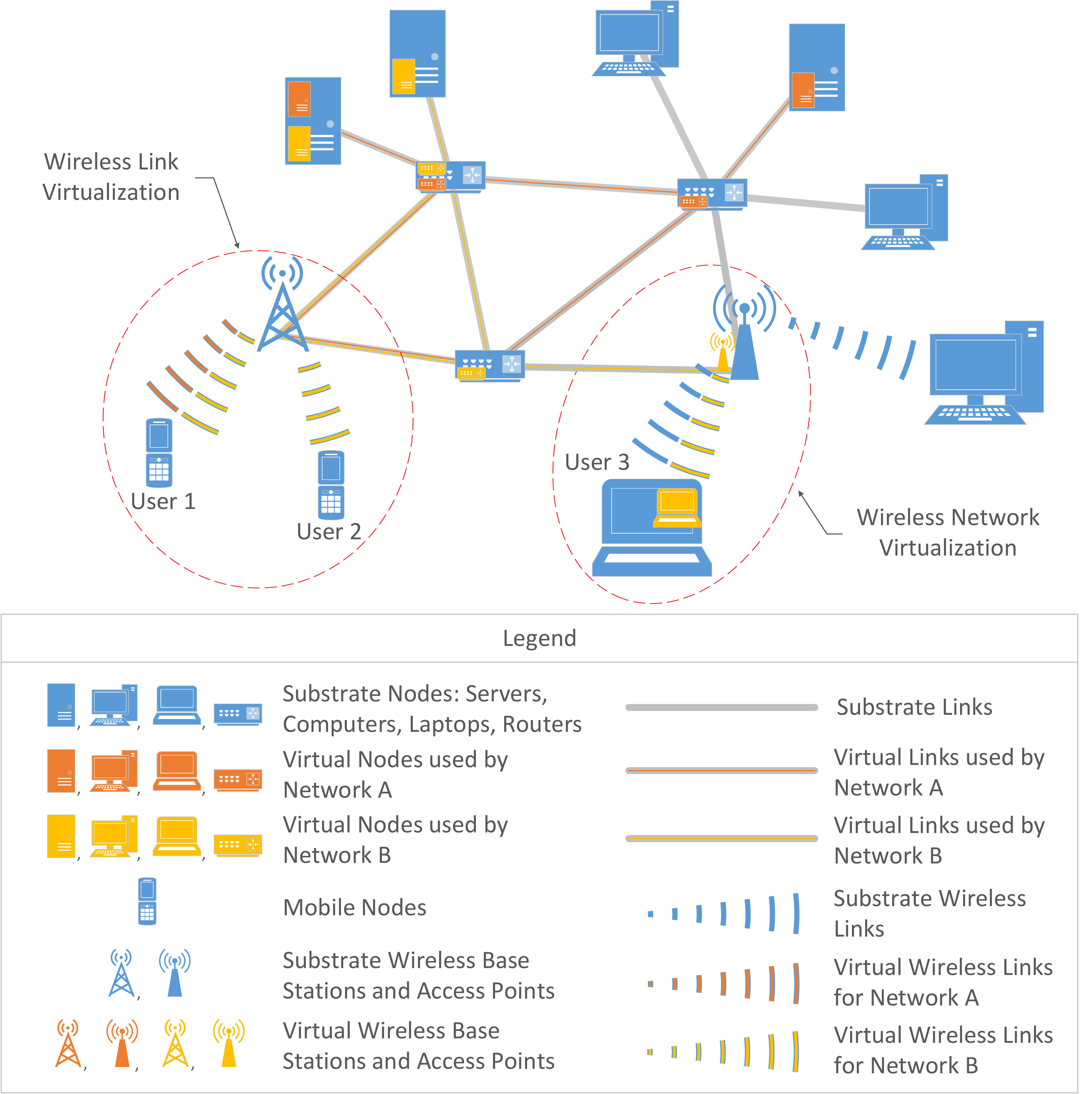}\hfill
  \caption[Wireless Link and Wireless Network Virtualization]{Wireless Link and Wireless Network Virtualization: Networks A (orange) and B (yellow) use network virtualization to coexist simultaneously on the wired network substrate. Wireless link virtualization is used to offer secure and customised services to users 1 and 2, and these services can focus on different aspects such as high-throughput, low-latency, or low-power consumption, etc. In this example, user 1 has access to two types of services, one from networks A and B each, while user 2 only has access to the service from network B. Wireless network virtualization is used to extend virtual network B to user 3.}
  \label{fig:wnv}
\end{figure}

Similarly to network virtualization, we can think of WNV as using both node and wireless link virtualization to create virtual wireless networks. Only when it is possible to virtualize both the node and link resources that make up the network, is it possible to consider (wireless) network virtualization \cite{Chowdhury2009} \cite{Wang2013Network}. As we have seen previously, node virtualization and wired link virtualization are long-established techniques. However, wireless link virtualization is still at an early stage of development, and without WLV it is not possible to do wireless network virtualization. In fact, most work on wireless network virtualization to date has actually been concerned with wireless link virtualization, since WLV is a prerequisite of WNV. In this work, we distinguish between wireless link virtualization and wireless network virtualization, shown in Figure \ref{fig:wnv}.


Although we are ultimately interested in wireless network virtualization, in this work we focus on wireless link virtualization for several reasons:

\begin{enumerate}
\item WLV is a prerequisite for WNV, and the other prerequisite, node virtualization, can already be implemented through many existing techniques.
\item The main difference between NV and WNV is the nature of the links, and thus once link virtualization has been accomplished, we could reuse NV techniques for other aspects of WNV, such as coordinated node and link mapping \cite{Chowdhury2009_2}.
\item Work on WNV so far has almost exclusively focused on WLV, although it has self-identified as WNV.
\end{enumerate}  

Now knowing that WLV is the main research aspect of WNV, in the next section we perform a survey of existing WLV techniques (although many of these techniques self-identify as wireless \textit{network} virtualization) in order to identify open research problems. We use the theory of virtualization to aid us with the classification and analysis of these works.

%
%

		\begin{table*}[tph]
		\caption{Non-Recursive Wireless Link Resource Allocation Techniques}
	   	\label{table:resource_allocation}
	    \centering
	     \resizebox{0.8\paperwidth}{!}{
	   	\begin{tabular}{ l l l l l c}
	   	
	 	\hline \\ [-1.5ex]
	 	& \textbf{First Author} & \textbf{Reference} & \textbf{Input Resource Type (Units)} & \textbf{Output Resource Type (Units)} & \textbf{Recursive?} \\ [0.5ex]
	 	\hline \\ [-2ex]
		\parbox[t]{4mm}{\multirow{32}{*}{\rotatebox[origin=c]{90}{\large Resource Allocation Techniques \hspace{8mm}}}}
		& Lu & \cite{Lu2012} & Subchannels (KHz) & Rate (Kbit/s)& No \\ [0.5ex]
		& Lv & \cite{Lv2012} & Channels (MHz) & Bandwidth -- i.e. rate (Mbit/s) & No \\ [0.5ex]
 		& Xu & \cite{Xu2013}, \cite{Xu2014} & Subchannels (MHz) & Sum Rate (bit/s/Hz) & No \\ [0.5ex]
 		& G. Liu & \cite{Liu2014}, \cite{Liu2015} & Subchannels (MHz) & Rate (Mbit/s) & No \\ [0.5ex]
 		& Kamel & \cite{Kamel2015} & Subchannel (MHz) & Rate (Mbit/s) & No \\ [0.5ex]
 		& G. Zhang & \cite{Zhang2015}, \cite{Zhang2015_2} & Subcarrier (MHz) & Data Rate (bit/s) & No \\ [0.5ex]
 		& Feng & \cite{Feng2015} - Spectrum & Spectrum (MHz) & Traffic (Mbit/s) & No \\ [0.5ex]
 		& Cai & \cite{Cai2015} & Radio Resource (KHz) & Utility (Bit/s/\$) & No \\ [0.5ex]
 		& Chen & \cite{Chen2015} & Channel (MHz) & Utility (Profit/bitrate)? & No \\ [0.5ex]
 		& Liang & \cite{Liang2015_2} & Radio Resource (MHz) & Utility (?) & No \\ [0.5ex]
 		& Khatibi & \cite{Khatibi2015} & Radio Resource Unit (MHz)? & Rate(Mbit/s) & No \\ [0.5ex]
 		& Rahman & \cite{Rahman2016} - LVN & Spectrum (MHz) & Rate (Mbit/s) & No \\ [0.5ex]
 		\cline{2-6} \\ [-2ex]
 		
 		& Fu & \cite{Fu2010}, \cite{Fu2012} & Fraction of time of subchannel (ms, KHz) & Information-theoretic Rate (Kbit/s) & No \\ [0.5ex]
 		& Kamel & \cite{Kamel2014} & Physical Resource Blocks (KHz, ms) & Rate (Mbit/s) & No \\ [0.5ex]
 		& Zaki & \cite{Zaki2011}, \cite{Zaki2011_2} - Dynamic & Physical Resource Blocks (KHz, ms) & Load (Kbit/s) & No \\ [0.5ex]
 		& Zhao & \cite{Zhao2011} & Physical Resource Blocks (KHz, ms) & Rate, Delay (Mbit/s, ms) & No \\ [0.5ex]
 		& Li & \cite{Li2012} & Physical Resource Blocks (KHz, ms) & Date Rate (Kbit/s) & No  \\ [0.5ex]	
 		& B. Liu & \cite{Liu2013} & Physical Resource Blocks (KHz, ms) & Traffic Rate (Kbit/s) & No \\ [0.5ex]
 		& Panchal & \cite{Panchal2013}  & Physical Resource Blocks (KHz, ms) & Load (Kbit/s) & No \\ [0.5ex]
 		& Kalil & \cite{Kalil2014} \cite{Kalil2016} & Resource Blocks (KHz, ms) & Throughput (Mbit/s) & No \\ [0.5ex]
 		& Moubayed & \cite{Moubayed2015} & Resource Blocks (KHz, ms)& Sum Rate (Kbit/s) & No \\ [0.5ex]
 		& Kokku & \cite{Kokku2012}, \cite{Kokku2013} & OFDM Blocks/Slots (MHz, ms) & Rate/QoS (Mbit/s) & No \\[0.5ex]
 		& Costa-Perez & \cite{Costa2013} & LTE Resource Blocks (MHz, ms) & Rate (Kbit/s) & No \\[0.5ex]
 		& Wei & \cite{Wei2015} & Physical Resource Blocks (KHz, ms) & Utility (bit/Joule) & No \\ [0.5ex]
 		& Wang & \cite{Wang2015} & Physical Resource Blocks (MHz, ms) & Throughput(Mbit/s) & No \\[0.5ex]

		\cline{2-6} \\ [-2ex]
		& Di Stasi & \cite{DiStasi2013} & Channels in Space (Hz, Grid Location) & Throughput (Mbit/s) & No \\ [0.5ex]
		& Q. Zhu & \cite{Zhu2015_2}, \cite{Zhu2015} & Subchannels and Power (Hz, Watts) & Utility (Hz/$\textrm{m}^3$)? & No \\ [0.5ex]
		& Fan & \cite{Fan2015} & Bandwidth and Power (MHz, dBm) & Rate (Mbit/s) & No \\ [0.5ex]
		& \multirow{2}{*}{K. Zhu} & \multirow{2}{*}{\cite{Zhu2016}} & \multirow{2}{14em}{Subchannels, Power and Antennas (Hz, Watts, \# of antennas)} &  \multirow{2}{*}{Rate (bits/s)}& \multirow{2}{*}{No} \\ [0.5ex]
		& \\
	 	& Ahmadi & \cite{Ahmadi2016} & Spectrum, Antennas (KHz, \# of Antennas) & Rate (Mbit/s) & No \\ [0.5ex]	 
		& Ahmadi & \cite{Ahmadi2016_2} & Spatial Stream (MHz, time block) & Rate (Mbit/s) & No \\ [0.5ex]
		& Rahman & \cite{Rahman2016} - CVN/RVN & Spectrum, Remote Radio Heads (MHz, \# of) & Rate (Mbit/s) & No \\ [0.5ex]
 		
 		\hline
	\end{tabular}
	}
\end{table*}

\section{Survey of Wireless Link Virtualization}
\label{sect:survey}

We now apply the theory of virtualization to the main focus of the paper; the survey on wireless link virtualization to establish open research questions.

\subsection{Test for Virtualization}
\label{subsect:test}

The first part of performing the survey was to develop a test for virtualization, to decide whether a technique that self-identifies as virtualization actually is virtualization. We return to the definition of virtualization to define such a test.

According to the definition of virtualization, virtualization is always recursive in theory. Therefore, one possible test for virtualization is to check for recursion. We propose a test which determines whether proposed techniques are recursive, at least in theory. 

All of the papers which we examine define abstract wireless link resources in some manner, and subsequently allocate these resources to different users based on some criteria. The allocation of resources to users is done through the use of an algorithm, whether this is an optimization, a heuristic method, an auction-based approach, or any other resource allocation method. These resource allocation methods all have one thing in common; they have a set of \textbf{input} resources and a set of \textbf{output} resources. Input resources are the wireless link resource that algorithms allocate, while output resources are the resources that the users receive from algorithms. In general, input resources are relatively straightforward to determine, but output resources can be harder to identify.

Therefore we base our test for recursion on the mapping of input to output resources. We define recursive techniques as techniques in which no difference exists between input and output resources. In other words, the \textit{types} of both resources are the same, and both the input and the output resources can be thought of as at the \textit{same} representation level. Techniques for which this is true can be considered recursive, and hence virtualization, and we can call the input and output resources real and virtual resources respectively.


The recursive test for virtualization is very simple:

\begin{mdframed}[
    linecolor=black,
    linewidth=0.5pt,
    roundcorner=0pt,
    backgroundcolor=white,
    userdefinedwidth=0.49\textwidth,
]
\centering 
\textbf{\normalsize Recursive Test for Virtualization} 
\end{mdframed}
\vspace{-5mm}
\begin{mdframed}[
    linecolor=black,
    linewidth=0.5pt,
    roundcorner=0pt,
    backgroundcolor=white,
    userdefinedwidth=0.49\textwidth,
]
\begin{enumerate}
\item Check the \textbf{units} of the input resources.
\item Check the \textbf{units} of the output resources.
\item If they are the same, then the technique can be considered recursive and is virtualization.
\end{enumerate}

\end{mdframed}

However, techniques for which the type of input and output resources differ cannot be considered recursive. In this case the input and output resources are at different representation levels. Users cannot use the same technique again to split or aggregate resources. These techniques can be considered methods of resources sharing/combination, but not virtualization. 

For example, a technique for which the input is spectrum resources (measured in Hertz) and the output given to users is also spectrum resources (also measured in Hertz), is considered a recursive technique, since users could use the same technique again to virtualize their own spectrum resources, and split or aggregate the resources in a new way.  

However, a technique which has as input spectrum resources (measured in Hz), and as output information rate (measured in bit/s) can not be considered recursive, since the units of the input and output resources are different. Users could not reuse this technique to split or aggregate their information rate. 

\subsection{Survey Overview}

The survey analyses over 60 works from 2003 onwards that self-identify as wireless network virtualization and propose algorithms for virtualization. However as discussed already, these works are in fact concerned with wireless link virtualization. Therefore the survey is about wireless link virtualization.

We apply the recursive test to a representative selection of papers which self-identified as virtualization, to determine whether these works were also deemed virtualization according to the formal definition of virtualization developed in this paper. We consider works on any type of wireless link technology, and at any layer of the communication stack. Only papers that propose resource mapping techniques are included, and thus many framework and architecture papers are omitted.

We split the papers into two groups: those that fail the recursive test, described in subsection \ref{subsect:fail}, and those which pass the recursive test and are therefore deemed to be relevant, analysed in subsection \ref{subsect:pass}. We briefly discuss why some papers did not pass the test; and why they are considered resource allocation techniques according to our definition of virtualization, rather than virtualization techniques.

\subsection{Non-Recursive Techniques}
\label{subsect:fail}

We briefly discuss the papers that did not pass the test, shown in Table \ref{table:resource_allocation}, to examine why they failed the test. Although these works are not virtualization according to our definition, these techniques still represent significant contributions in wireless link resource allocation and allow wireless link resources to be shared in an efficient and/or fair manner.




The works in Table \ref{table:resource_allocation} are organised by the type of resources that they allocate and by date. As can be seen, we found several types of papers which consider spectrum resources as input, whether isolated in frequency, frequency and time, or space and other dimensions. 

\subsubsection{Frequency-based isolation}
As an example of the recursive test, we look at the works \cite{Lu2012, Lv2012, Xu2013, Xu2014, Liu2014, Liu2015, Kamel2015, Zhang2015, Zhang2015_2, Khatibi2015, Rahman2016}. In these works, input spectrum resources are allocated as subchannels (in units of Hz) to users based on the rate (in bit/s) that can be achieved by users. Thus the output users are receiving is a data rate, rather than a number of subchannels. Although these methods advance the field of spectrum resource allocation, because the input and the output resources are not of the same type, these techniques cannot be considered recursive, and subsequently do not fall under our strict definition of virtualization. Similarly, the work \cite{Feng2015} proposes a spectrum resource sharing technique that relies on estimating the users' traffic requirements, and thus the output resources are at a higher level of representation than the input resources.

\subsubsection{Frequency and Time based isolation}
There are also techniques which allocate input spectrum resources using frequency and time isolation, either based on users' estimated traffic loads and requirements, or based on the best rates that users can achieve. The works \cite{Zhao2011, Li2012, Panchal2013, Liu2013, Kalil2014, Zaki2011, Zaki2011_2} can be considered in the first category, and take into consideration users' traffic requirements when assigning spectrum blocks. The second category contains the works \cite{Fu2010, Fu2012, Kamel2014}. Similarly, the works \cite{Kokku2012}, \cite{Kokku2013}, and \cite{Costa2013} implement resource allocation schemes that allocate wireless resources based on the data rate requested by the virtual networks. These implementations are very impressive resource sharing techniques, but again because the techniques are not recursive, they cannot be considered virtualization. 

\subsubsection{Space-based isolation}
\label{subsubsect:space}
Several works such as \cite{DiStasi2013, Zhu2015_2, Fan2015, Zhu2015, Zhu2016, Ahmadi2016, Ahmadi2016_2} propose allocating virtual spectrum resources using space-based isolation methods such as power control, and Multiple Input Multiple Output (MIMO) schemes. However, these schemes rely on the Channel State Information (CSI) to allocate virtual spectrum resources. Since CSI is an information-level resource, these schemes cannot be considered recursive. 

Moreover space-based isolation can only be achieved if the resource users do not care which spatial resources they receive - in other words if the location of the users does not matter. The reason for this is that resources that are requested and allocated in a particular direction are different to resources that have no spacial restrictions. For example, allocating bandwidth to a spatial stream means that units of the resource are now Hz in a particular direction of a coordinate system, and we can think of the spectrum resource as changing from a scalar to a vector resource. 

In fact, directed space-based isolation schemes enable spatial densification, and MIMO techniques can be seen as analogous to increasing the number of wireless nodes \cite{Bhushan2014}. Taking this perspective clarifies why directed spectrum resources are not the same as virtual spectrum resources.


Although the techniques discussed in this subsection provide clever and efficient methods to share limited spectrum resources efficiently, they take as input spectrum resources, but the output resources that users receive are at the information representation level (see next section). Although this leads to better spectral efficiency and more optimal use of resources, this means that these techniques cannot be considered virtualization. 


\subsection{Recursive Techniques}
\label{subsect:pass}

The papers which pass the recursive test are shown in Table \ref{table:survey}, where we have combined the input and the output resources types for brevity (since they are the same). We discovered that the recursive test could also be used to determine the representation level of a virtualization technique. Methods which have the same unit types (for example, Hz, KHz, MHz, etc.) can be grouped together at the same representation level. We found that there were two main representation levels for wireless link resources: 

\begin{enumerate}
\item Spectrum Level Virtualization, which the virtualization of wireless resources in the form of spectrum, such as \cite{Tan2012}.
\item Information Level Virtualization, also known as data path or flow-level virtualization, which refers to the virtualization of the information carried by the spectrum, and allows wireless links to be shared or combined at the information level, such as \cite{Chandra2004}.
\end{enumerate}


Table \ref{table:survey} classifies papers firstly according to the representation level, which is in our opinion the most important distinction between virtualization techniques. Subsection \ref{subsect:spectrum} discusses spectrum-level virtualization techniques, while subsection \ref{subsect:information} discusses information-level techniques.  We also examined the types of isolation used at each representation level, and in particular the isolation methods that seemed to be more successful. Therefore in the table, the second classification is by isolation dimension(s), ordered by date. The third consideration of the survey is the embedding method used by virtualization techniques. Due to space limitations, the different embedding methods are described in Table \ref{table:embedding}. A final aspect that we examined was the type of mapping that techniques performed, which was very heavily focussed on many-to-one mapping, i.e. partitioning. 

In fact, from the survey it became clear that several works have also touched on the concept of `levels of virtualization' or `depth of virtualization', such as \cite{Kokku2012, Yang2013, Wang2013Network, Liang2014, Liang2015, Wen2013, Wen2013Current}. However the majority of these works describe the levels in terms of the types of resources used, rather than different representation levels. From analysis of these works, it became clear that the terms `levels of virtualization' and `depth of virtualization' often refer to the concept of virtualization at different levels of representation, but that this is not articulated clearly in these works.

\subsection{Spectrum Level Virtualization}
\label{subsect:spectrum}

To the best of our knowledge, spectrum-level virtualization was first proposed in \cite{Paul2006}, in which wireless network virtualization is considered as an extension to the wired network virtualization testbed GENI. The authors of \cite{Paul2006} conceptually discuss several methods of isolation at the spectrum level, such as along the frequency, time, space, and code dimensions or a combination of these. The authors of \cite{Sachs2008} propose a virtualization framework for wireless networks and discuss some of the steps involved in deploying virtual wireless networks. 

		\begin{table*}[tph]
		\caption{Wireless Link Virtualization: Isolation}
	   	\label{table:survey}
	    \centering
	   \resizebox{0.82\paperwidth}{!}{
	   	\begin{tabular}{ l l l l c c l c }
	   	
	 	\hline \\ [-1.8ex]
	 	& \multirow{2}{7em}{\textbf{First Author }} & \multirow{2}{*}{\textbf{Reference}} & \multirow{2}{12em}{\textbf{Input \& Output \\ Resource Type (Units)}} & \multirow{2}{*}{\textbf{Recursive?}} & \multirow{2}{7em}{\centering \textbf{Representation Level}} &  \multirow{2}{9em}{\textbf{\hfill Isolation \hfill }} & \multirow{2}{*}{\textbf{Mapping}} \\ 
	 	& \\ [0.5ex]
	 	\hline \\ [-1.5ex]
	 	\parbox[t]{4mm}{\multirow{28}{*}{\rotatebox[origin=c]{90}{\Large Virtualization \hspace{2mm}}}} 
 		& Forde & \cite{Forde2011} & Bands (MHz) & Yes & \multirow{12}{1em}{\VastT\}} \multirow{12}{6em}{Spectrum} &  \multirow{3}{1em}{\vast\}} \multirow{3}{6em}{Frequency} & $m \mapsto 1$\\ [0.5ex]
 		& Tan & \cite{Tan2012} & Baseband (MHz) & Yes & & & \hspace{0.4em}$m \mapsto m$\\ [0.5ex]
	 	& Yang & \cite{Yang2013_2}, \cite{Yang2014_2} & Channel (MHz) & Yes & & & $m \mapsto 1$ \\ [0.5ex]
 		\cline{7-8} \\ [-1.8ex]
	 	& Zaki & \cite{Zaki2011}, \cite{Zaki2011_2} - Static & Physical Resource Blocks (KHz, ms) & Yes & & \multirow{5}{1em}{\vAstt\}} \multirow{5}{6em}{Frequency and Time} & $m \mapsto 1$\\ [0.5ex]
 		& Yang & \cite{Yang2012} & Physical Resource Blocks (KHz, ms) & Yes & & & $m \mapsto 1$\\ [0.5ex]	
 		& van de Belt & \cite{vandeBelt2014},  \cite{vandeBelt2015} & Physical Resource Blocks (KHz, ms) & Yes & & & $m \mapsto 1$\\ [0.5ex]
 		& Hsu & \cite{Hsu2015} & Bandwidth and Time units (KHz, ms) & Yes & & & \hspace{0.4em}$m \mapsto m$\\ [0.5ex]
 		& Lu & \cite{Lu2015} & Physical Resource Blocks (KHz, ms) & Yes & & & $m \mapsto 1$\\ [0.5ex]
		\cline{7-8} \\ [-1.8ex]
 		
 		& \multirow{2}{*}{X Zhang} & \multirow{2}{*}{\cite{Zhang2012}} & \multirow{2}{16.5em}{Timeslot and (x, y) coordinates in Space (ms, (meter, meter) from \cite{RaychaudhuriSeskarOttEtAl2005})} & \multirow{2}{*}{Yes} & & \hspace{0.7em} \multirow{2}{7em}{ Time and \\ \hspace{0.1em} Space - Power} & \multirow{2}{*}{$m \mapsto 1$}\\ [-2ex]

 		&\\ 
		& \\ 
 		\cline{2-8} \\ [-1.8ex] 
 		& Chandra & \cite{Chandra2004} & Information Throughput (Kbit/s) & Yes & \multirow{18}{1em}{\VAstT\}} \multirow{18}{6em}{Information} & \multirow{5}{1em}{\vAstt\}} \multirow{5}{6em}{Time - Switching} & $m \mapsto 1$	\\ 	[0.5ex]
 		& Smith & \cite{Smith2007} & Throughput (Kbit/s) & Yes & & &  $m \mapsto 1$ \\ [0.5ex] 
 		& Coskun & \cite{Coskun2009} & Throughput (Kbit/s) & Yes & & &  $m \mapsto 1$ \\ [0.5ex]
 		& AlHazmi & \cite{AlHazmi2011} & Packets (Bytes) & Yes & & & $m \mapsto 1$ \\ [0.5ex]
 		& Lv & \cite{Lv2012_2} & Packets (Bytes) & Yes & & & $m \mapsto 1$ \\ [0.5ex]
 		 \cline{7-8} \\ [-1.8ex]		 			
	 			
 		& Mahindra & \cite{Mahindra2008} & Throughput (Mbit/s) & Yes & & \multirow{10}{1em}{\vspace{1mm}\VaSt\}} \multirow{10}{6em}{Time - Addressing} \vspace{-1mm} &  $m \mapsto 1$ \\ [0.5ex]
 		& Perez & \cite{Perez2009} & Packets (Bytes) & Yes & & &  $m \mapsto 1$ \\ [0.5ex]
 		& Sherwood & \cite{Sherwood2009} & Throughput (MBit/s) & Yes & & &  $m \mapsto 1$ \\ [0.5ex]
 		& Bhanage & \cite{Bhanage2010}, \cite{Bhanage2010_2} & Throughput (Mbit/s) & Yes & & & $m \mapsto 1$ \\ [0.5ex]
 		& Xia & \cite{Xia2011} & Throughput (Kbit/s) & Yes & & &  $m \mapsto 1$ \\ [0.5ex]
 		& Aljabari & \cite{Aljabari2011} & Throughput (Kbit/s) & Yes & & &  $m \mapsto 1$ \\ [0.5ex]
 		& Nakauchi & \cite{Nakauchi2012} & Throughput (Mbit/s) & Yes & & & $m \mapsto 1$ \\ [0.5ex]
 		& Katsalis & \cite{Katsalis2014} & Throughput (Mbit/s) & Yes & & & $m \mapsto 1$ \\ [0.5ex]
 		& Feng & \cite{Feng2015} - Flow & Rate(Mbit/s) & Yes & & & $m \mapsto 1$ \\ [0.5ex]
 		\cline{7-8} \\ [-1.8ex] 	
 			
 		& Mahindra & \cite{Mahindra2008} & Throughput (Mbit/s) & Yes &  & \multirow{1}{8em}{\hspace{1.2em} Space - Power } &  $m \mapsto 1$ \\ [1ex] 	

 		\hline \\ [-2ex]
	\end{tabular}
	}
\end{table*}        
        

\subsubsection{Frequency-based Isolation}
\label{subsubsect:frequency}

One of the first proposals of a spectrum virtualization technique was \cite{Forde2011}. This work proposed an auction-based method for mapping virtual spectrum bands to real spectrum bands, and simulated the allocation of resources to maximize the revenue for the auctioneer. 

The work \cite{Tan2012} implements a Spectrum Virtualization Layer (SVL) that offers virtual spectrum resources to users according to their desired bandwidth. In this way the limitations of real spectrum with fixed bandwidths and non-contiguous segments can be overcome. The virtual spectrum is mapped to real spectrum according to a predefined spectrum map. The wireless link mapping is many-to-many as it can map the virtual baseband to a real spectrum band with the same bandwidth, a narrower bandwidth, or to several non-contiguous bands with a greater combined bandwidth. The authors show that SVL can be used to create a virtual Wifi Network over TV White Space channels, and that SVL could be used to combine the access points of several different technology standards.

The authors of \cite{Yang2013_2} and \cite{Yang2014_2} propose a channel virtualization scheme to share spectrum opportunistically, based on the price that virtual networks are willing to pay for virtual channels. The requirements of virtual networks are split into baseline and fluctuant parts. In this work the virtual channels used for the fluctuant requirements can be shared between different virtual networks and thus there is a probability of collision. The authors propose several algorithms to assign the virtual resources in such a way that maximizes the revenue for the owner of the real spectrum, and perform simulations to compare these and other algorithms from wired network virtualization. 

\subsubsection{Frequency and Time based Isolation}

The works \cite{Zaki2011_2} and \cite{Zaki2011} first propose a spectrum virtualization method involving frequency and time isolation. The idea in these works is to virtualize Long Term Evolution (LTE) base stations, also called eNodeBs, using a virtual machine hypervisor that also has control of the spectrum resources. The virtual eNodeBs can request physical radio resource blocks (PRBs) from the hypervisor, The authors propose two different algorithms - a static algorithm where the virtual PRBs are assigned once, and a dynamic algorithm which depends on the operator's load. However, only the static algorithm can be considered virtualization, because the dynamic algorithm fails the test for recursion since the hypervisor assigns virtual PRBs based on a higher level of representation (information-level metrics such as channel conditions, user loads and QoS requirements).

The authors of \cite{Yang2012} propose a virtual resource embedding algorithm for frequency and time domain resources. In this work, virtual network operators can request a number of contiguous frequency and time-domain resources for a duration of time periods. Virtual resource requests can be chosen with several priority levels, corresponding to the likelihood of acceptance of a request. The heuristic algorithm proposed is based on finding the most suitable location for a request using Karnaugh-map theory and the embedding density, a measure of how efficiently resources are used. The authors compare several algorithms through simulations in terms of resource utility and the acceptance ratio of virtual resource requests.

Our work \cite{vandeBelt2014} extends the Karnaugh-map algorithm to the dynamic case, which allows virtual resources to be re-embedded at every time period. Through simulations, we show that this uses resources in a more efficient manner and leads to a higher request acceptance ratio. We also propose a greedy dynamic embedding algorithm which has increased performance compared to the dynamic Karnaugh-map algorithm, and formulate the objective problem for the virtual resource embedding problem with prioritised requests. 

In another work, \cite{vandeBelt2015}, we demonstrate how virtualization could be applied to a use-case scenario, in the context of a virtual public safety operator. We show that virtualization enables spectrum resources to be used in a flexible manner that guarantees connectivity for public safety users during an emergency, but allows spectrum resources to be used by commercial operators as needed during day-to-day operation. 

The work \cite{Hsu2015} proposes a dynamic virtual resource embedding algorithm that allows virtual resource requests to be fulfilled through multiple virtual resource blocks. In other words the real resource owner can aggregate spectrum blocks to fulfil resource requests. This use of spectrum aggregation means that the embedding used is potentially many-to-many. 

In the work \cite{Lu2015}, the dynamic greedy algorithm is extended by using a genetic algorithm to find the most fit embedding location for a virtual resource request. The authors show that the genetic algorithm offers a small improvement over the dynamic greedy algorithm in terms of resource utility and rejection rate. 

\subsubsection{Time and Space based Isolation}

The work \cite{Zhang2012} proposes a time and space based isolation method to allocate wireless network resources to experiments on a grid. The objective is to fit as many experiments into a limited grid, while maintaining strict isolation. The allocation takes into account the distance between nodes as a means of modelling potential interference. The authors show that using space-based isolation in addition to time-based isolation can provide significant gains in resource utilization. Gains are more significant for small scale experiments than for larger experiments.
One important aspect to note is that space-based isolation is possible in this case because the location of the nodes does not matter. However, in more commercial applications this assumption most likely will not be true, and therefore space-based isolation will not be possible (see section \ref{subsubsect:space}).


\begin{table*}[tph]
		\caption{Wireless Link Virtualization: Embedding}
	   	\label{table:embedding}
	    \centering
	   \resizebox{0.8\paperwidth}{!}{
	   	\begin{tabular}{ l l l c l }
	   	
	 	\hline \\ [-1.8ex]
	 	& \textbf{First Author} & \textbf{Reference} & \textbf{\hfill Isolation \hfill} & \textbf{Embedding} \\ [0.5ex]
	 	\hline \\ [-1.8ex]
	 	\parbox[t]{4mm}{\multirow{13}{*}{\rotatebox[origin=c]{90}{\large Spectrum Level \hspace{1mm}}}} 
	 	
 		& Forde & \cite{Forde2011} & \multirow{4}{1em}{\Vast\}} \multirow{4}{6em}{Frequency} & Auctioning of paired spectrum bands based on price and distance between bands \\ [0.5ex]
 		& Tan & \cite{Tan2012} & & Implementation of spectrum virtualization based on predefined spectrum map(s) \\ [0.5ex]
	 	& \multirow{2}{*}{Yang} & \multirow{2}{*}{\cite{Yang2013_2}, \cite{Yang2014_2}} & & \multirow{2}{31em}{Baseline and varying channels with collision probability. Comparison of heuristic, genetic, and optimal allocation algorithms based on price}  \\ 
	 	
	 	& \\ 
 		\cline{4-5} \\ [-1.8ex]
	 	& Zaki & \cite{Zaki2011}, \cite{Zaki2011_2} - Static &  \multirow{5}{1em}{\vAstt\}} \multirow{5}{6em}{Frequency and Time} & Predefined spectrum allocation \\ [0.5ex]
 		& Yang & \cite{Yang2012} & & Heuristic embedding based on Karnaugh-map areas  \\ [0.5ex]
 		& van de Belt & \cite{vandeBelt2014},  \cite{vandeBelt2015} & &  Dynamic greedy embedding and dynamic version of \cite{Yang2012} \\ [0.5ex]
 		& Hsu & \cite{Hsu2015} & & Bottom-Left algorithm with splitting of resources \\ [0.5ex] 
 		& Lu & \cite{Lu2015} & & Dynamic genetic algorithm for Karnaugh-map areas  \\ [0.5ex]
		\cline{4-5} \\ [-1.8ex]
 		& \multirow{2}{*}{X Zhang} & \multirow{2}{*}{\cite{Zhang2012}} & \hspace{1.2em} \multirow{2}{7em}{ Time and \\ \hspace{0.1em} Space - Power} & \multirow{2}{*}{Embedding based on distance between nodes and experiment duration} \\ [0.5ex] 		
 		&\\
 		\cline{2-5} \\ [-1.8ex] 	
 		
 		\parbox[t]{4mm}{\multirow{15}{*}{\rotatebox[origin=c]{90}{\large Information Level \hspace{16mm}}}} 
 		& Chandra & \cite{Chandra2004} & \multirow{5}{1em}{\vAstt\}} \multirow{5}{6em}{Time - Switching} & Time-switching at either fixed intervals or adaptive intervals based on traffic load \\ [0.5ex]
 		& Smith & \cite{Smith2007} & & Synchronised time-switching in a round-robin fashion  \\ [0.5ex]
 		& Coskun & \cite{Coskun2009} & & Fast time-switching at regular intervals using PSM and PCF features \\ [0.5ex] 
 		& AlHazmi & \cite{AlHazmi2011} & & Extension of \cite{Coskun2009} with more detailed analysis of virtualization capabilities \\ [0.5ex]
 		& Lv & \cite{Lv2012_2} & & Round-robin time-switching with opportunistic rebroadcasts \\ [0.5ex]
 		\cline{4-5} \\ [-1.8ex]		 			
	 			
 		& Mahindra & \cite{Mahindra2008} & \multirow{11}{1em}{\VasTt\}} \multirow{11}{6em}{Time - Addressing} & Policy manager limits each experiment to a predefined maximum rate \\ [0.5ex]
 		& Perez & \cite{Perez2009} & & Weighted round-robin scheduler \\ [0.5ex]
 		& Sherwood & \cite{Sherwood2009} & & Priority resource allocation policy configured by network administrator \\ [0.5ex]
 		& Bhanage & \cite{Bhanage2010} & & Based on airtime quotas allocated at hardware setup.  \\ [0.5ex]
 		& Bhanage & \cite{Bhanage2010_2} & & Weighted sharing based on an airtime fairness metric \\ [0.5ex]
 		& Xia & \cite{Xia2011} & & Not clear how vMAC interfaces are embedded but possibly through rate control \\ [0.5ex] 
 		& Aljabari & \cite{Aljabari2011} & & No specific embedding - limited by machine performance \\ [0.5ex]
 		& Nakauchi & \cite{Nakauchi2012} & & Embedding based on a weighted fraction of airtime \\ [0.5ex]
 		& Katsalis & \cite{Katsalis2014} & & Allocated according to predefined throughput share guarantees \\ [0.5ex]
 		& Feng & \cite{Feng2015} - Flow & & Flow scheduler allocates flows to virtual networks based on traffic load\\ [0.5ex]
 		\cline{4-5} \\ [-1.8ex] 	
 			
 		& Mahindra & \cite{Mahindra2008} & \hspace{1.2em} Space - Power  & Policy manager limits each experiment to a predefined maximum rate \\ [0.5ex] 	

 		\hline \\ [-1.8ex]
		\end{tabular}
	}
\end{table*}  

\subsubsection{Discussion}

Spectrum-level virtualization has so far mainly focused on frequency isolation, plus the combination of isolation by frequency and other dimensions such as time or space. One reason for this might be that frequency isolation is already extensively used in networks as frequency division multiplexing. 

There are several reasons that could explain why frequency isolation is so widely used. The first is that due to the nature of spectrum resources, different parts of the spectrum are suitable for different applications and thus it is intuitive to divide up the spectrum into bands for different applications. Another reason why frequency isolation could be preferable to time isolation is that it allows wireless links to be operational at all times, whereas using time isolation this is not possible. Time isolation also requires that nodes in a network be synchronised tightly. 

Space isolation is also used widely in wireless network and allows the spectrum to be reused in multiple locations. However space isolation is possibly the most difficult form of ensuring independence between different links, due the unreliable and random nature of propagation of radio waves.

With the exception of \cite{Tan2012}, to date the work on spectrum virtualization has focussed heavily on simulation and on efficient embedding schemes, rather than on implementing virtualization techniques with strict isolation. One of the reasons for this might be that most of the works deal with cellular technologies, and thus there is an emphasis on efficiency, as it is costly and time-consuming to add additional capacity to these types of networks. Another reason could be the challenges of implementation of virtualization in these networks, since many techniques for cellular networks rely on strict synchronisation.

\subsection{Information Level Virtualization}
\label{subsect:information}

The concept of wireless network virtualization was first seen as an extension to wired network virtualization \cite{Paul2006}, \cite{Spyropoulos2007}, and initial wireless link virtualization techniques focused on information-level link virtualization, which had already been accomplished for wired networks (see section \ref{subsect:VPNs}). The idea of Virtual Local Area Networks (VLANs) had also been established for wired networks, and so some of the first wireless link virtualization papers focused on finding methods to allow multiple virtual links to coexist on the same physical wireless device at the information level.

\subsubsection{Time-based Isolation using Switching}

One of the first papers to describe and implement a method of wireless link virtualization was \cite{Chandra2004}. In this paper, the authors address the problem of connecting a wireless node to multiple networks simultaneously, which up until this point could only be done using multiple wireless front-ends. To solve the problem of expense and excessive battery consumption by multiple wireless cards amongst other reasons, the MultiNet solution was developed, which time-switched a wireless card between multiple virtual networks at regular intervals, or at adaptive intervals based on traffic load. Virtual interfaces are presented to the node at the information level (Media Access Control (MAC) layer), and the node can send and receive packets through one or more interfaces simultaneously. The interfaces appear as if they are constantly active, but in practise the device driver buffers packets until the right virtual network is active. This solution reduced the energy consumed dramatically compared to using multiple cards, since two radios consumed about double the energy consumed by MultiNet. However, the process of switching between networks means that the association procedure is initiated every time a switch occurs, and leads to a decrease in performance in terms of delay, throughput and packet loss when compared to solutions involving multiple wireless cards. 

The work \cite{Smith2007} proposes a round-robin time-switching approach for virtualizing wireless links used by virtual nodes. These virtual networks are co-ordinated centrally to ensure synchronization between nodes. However, because of propagation delay, some overhead is incurred when switching from one virtual network to another and this leads to a relatively high packet loss and increased latency.

To deal with the issues of long handover times when switching between virtual links, the authors of \cite{Coskun2009} use the Power Saving Mode (PSM) to shorten delays compared to virtualization without PSM enabled. They also take advantage of the Point Coordination Function (PCF) features of wireless cards to avoid repeating the association procedure for every switch. In \cite{AlHazmi2011}, an extension of \cite{Coskun2009}, the authors investigate the effects of virtualization on traffic properties.

The work \cite{Lv2012_2} examines time-switching in the context of wireless mesh networks, and proposes a round-robin scheme to broadcast packets from differing virtual networks. Because links can be unreliable, and no acknowledgement is used when using the broadcast mode, the algorithm rebroadcasts certain packets based on a successful reception probability.

However some of the disadvantages of the time-switching approach are that a high level of synchronization is required between different nodes, and also that when there are many virtual networks the delay for each network increases, which affects the performance of delay-sensitive applications. 

\subsubsection{Time-based Isolation using Addressing}

The concept of Virtual Access Points (VAPs) was first introduced in \cite{Aboba2003} as a means of allowing multiple access points (and thus multiple networks) to coexist simultaneously on one physical access point. Each VAP is allocated a unique Service Set Identifier (SSID) and capability set. The use of addressing in this way provides time isolation at the information level (MAC layer) and enables VAPs to be indistinguishable from physical access points \cite{Aboba2003}. VAPs cannot emulate the operation of a physical AP at the radio frequency layer (Spectrum level) unless multiple physical radios are available. This limitation requires all virtual networks to use the same physical radio parameters such as channel. However, VAPs can differentiate at the information level, offering different throughput rates, packet loss, and latency, etc.

One of the first papers to implement Virtual Access Points is \cite{Mahindra2008}. This paper compares space-based and time-based isolation for virtualization on the ORBIT testbed. Space-based isolation will be discussed later. For the time-based isolation the authors prefer the VAP approach rather than the time-switching approach, because of the disadvantages of the time-switching approach already discussed. A policy manager limits the maximum rate that each experiment can achieve. The authors show that using VAPs adds minimal overhead to the throughput performance and the delay compared to the conventional approach.

The work \cite{Perez2009} investigates a virtualization approach that stores packets from different virtual operators in separate queues until the scheduler sends packets to the wireless interface using a weighted round-robin scheme. Although this approach does not use VAPs, it also uses packet addresses to isolate at the information level. 

Similarly, in \cite{Sherwood2009}, addressing-based isolation is used to partition the network into different `slices', which can have different data rates set by the network administrator. Address-based isolation is also used in \cite{Bhanage2010_2} to create a virtual WIMAX base-station that can offer slice customization to its users. The resource allocation is based on airtime quotas configured at setup.

A method for controlling the air-time usage of different virtual networks is proposed in \cite{Bhanage2010} and \cite{Nakauchi2012}. The virtual networks are isolated using virtual access points and the slice identifiers are used to monitor and control air-time usage of each network. 

Virtual WiFi \cite{Xia2011} proposes a method of having multiple MAC interfaces for virtual machines on the same physical wireless device. This is achieved through giving a different address to each virtual MAC (vMAC) interface and assigning each virtual machine a MAC address. However, it is not clear whether vMACs have any resource limits (apart from the hard performance limits of the device).

The authors of \cite{Aljabari2011} also use VAPs to deploy multiple wireless networks on a shared physical infrastructure. The focus is on open-source virtualization of access points and wireless network interfaces. However, the embedding used is not described. 

In \cite{Nakauchi2012} virtual networks are isolated through MAC addressing, and each virtual network has its own queue with different parameters. The algorithm to assign wireless airtime to virtual networks is based on a weighted fraction. The authors show that the desired fractions of airtime can be achieved quite closely using their resource allocation algorithm.

The CONTENT project aims to investigate end-to-end virtualization in heterogeneous wireless and optical networks. In the project, wireless virtualization is done by assigning each flow a virtual identification tag, similar to the above approaches \cite{Katsalis2014}. The embedding is performed according to predefined throughput guarantees.

The work \cite{Feng2015} developed two algorithms, one a spectrum-level algorithm and the other a flow/information level technique that uses addressing to allocate virtual packets based on traffic load. However only the flow level technique can be considered virtualization, since the spectrum level technique allocates spectrum also based on traffic load, and thus is not recursive.

\subsubsection{Space-based Isolation}

As mentioned previously, the authors of \cite{Mahindra2008} also examine space-based isolation for virtual networks. Isolation is achieved between nodes on a shared channel by using sufficient spatial separation to avoid interference and by managing the power levels of transmissions. Crucially, this work is concerned with allocating experiments on a grid, and space-based isolation is appropriate in this scenario, since it does not matter \textit{where} experiments are located in space.

\subsubsection{Discussion}

Although time-switching was initially found to be a potential solution for wireless link virtualization at the information level, it soon became apparent that the delays between switching led to significant performance decreases in terms of throughput and especially latency. Another problem with time-switching is the need for precise synchronization between different nodes. The addressing-based isolation method provides many advantages in this regard compared to time-switching. 

However, the addressing-based solution could have some overheads compared to time-switching since every node receives all of the traffic, and packets are only filtered out at the receiving nodes. This could lead to reduced battery life, an important point to consider for wireless nodes.

The space-based approach does not seem to offer many advantages compared to the time-switching or the addressing-based approaches, since isolation between different virtual links is very difficult to achieve reliably, and thus this affects the performance of the virtual links.

In terms of the embedding techniques used, the majority of works employ either predefined resource allocation, or round-robin style allocation. Only \cite{Chandra2004} and \cite{Feng2015} propose embedding schemes based on traffic load. It is clear that the focus of information level techniques is on achieving isolation and actually implementing virtualization. A possible explanation of the emphasis on isolation is that these virtualization schemes are used in scenarios where the resource capacity can easily be increased, such as WiFi. 

\subsection{Comparison with Existing Surveys}
\label{subsect:lit_rev}

As previously mentioned, there are several surveys concerning wireless link and network virtualization, amongst which the most important are \cite{Wen2013Current, Liang2014, Wen2013, Yang2014, Richart2016}. We briefly compare the approaches taken by these works to this paper, and highlight how this paper offers additional and complementary insights to the existing literature. Typically such a review would come before performing the survey; however, in this case, now that the survey is complete, it is easier to show how the approach taken in this paper provides additional understanding. 

\subsubsection{Focus on Purpose / Technology}

In general, existing surveys define virtualization in terms of its perceived purpose (e.g. in terms of business models, benefits etc.), and classify virtualization techniques by purpose and technology. For example in \cite{Richart2016} wireless network virtualization techniques are grouped by access technology, such as IEEE 802.11 (WiFi), 3GPP LTE, IEEE 802.16 (WiMAX), while in \cite{Wen2013}, \cite{Yang2014}, and \cite{Liang2014} techniques are grouped by technology and purpose -- whether techniques have been developed for commercial or experimental use.

Although the above approaches are useful for giving a detailed description of work that has been done to date, it can be hard to identify universal virtualization trends and potential research topics. By taking a purposely technology-agnostic approach based on a strong definition of virtualization, in this work we are able to classify virtualization techniques theoretically. The advantage of this is that it is easier to identify common trends (described in the next section) and propose problematic research areas in a clear and coherent manner. In addition, the technology-agnostic approach has another benefit; it can still be relevant to future technologies when these are developed.

As a specific example, compare the above Table \ref{table:survey} with Table I found in \cite{Richart2016}. Both tables cover a similar collection of papers (some papers found in \cite{Richart2016} do not pass the recursive test, and are omitted), but in the latter table techniques are grouped by technology, whereas in this work techniques are grouped by representation level. In the technology-oriented table, all of the works on LTE (\cite{Zaki2011, Zaki2011_2, Yang2012, vandeBelt2014}) can be found at the spectrum-level in the other table, and the works on 802.11 (\cite{Nakauchi2012, Smith2007, Mahindra2008, Xia2011}) can be found at the information-level. By focusing on technology, one might wrongly assume that virtualization in 802.11 can only be done through information-level techniques such as traffic shaping, or that virtualization in LTE can only be achieved by spectrum virtualization. However, virtualization can always be done at any representation level; for example, 802.11 virtualization occurs at the spectrum-level in \cite{Tan2012}. The \textit{observation} that to date most work in LTE has been at spectrum-level, and most work in WiFi has been at information-level is only possible because of the technology-agnostic approach.

Thus the focus on the purpose/technology of virtualization techniques offers certain insights and enables the detailed description of virtualization techniques, but it can mean that general trends and new possibilities are missed.

\subsubsection{Lack of Specificity}

The theoretical approach underpinning this survey provides a second advantage over existing surveys; it offers a specific language that is clear and succinct, which can lead to new insights about virtualization.

Several of the concepts described in this paper are briefly mentioned in the current literature. For example, the need for isolation in virtualization is mentioned in almost all surveys listed above. However, although isolation is recognised as important, it is not explicitly described. In \cite{Liang2014} and \cite{Richart2016} isolation is used for the survey tables, although a vast range of different isolation measures are given, among which are: MAC layer, time division/slot, PRBs, packet, spatial, flow, traffic shaping, rate control, slice, sub-channel/sub-carrier. The large number of isolation types does not provide clarity, but instead means that isolation is left vague. By using a theoretical approach in this survey, a clear understanding of isolation is reached, and we can identify the specific isolation types such as frequency, time, etc., even when different terms are used. 

Similarly, the concept of `virtualization levels' can be found in \cite{Liang2014, Richart2016, Wen2013} and many other works on virtualization. Again, this concept is insufficiently described and is unclear. Often `levels' of virtualization are described as spectrum-, infrastructure-, and network-level virtualization. In fact, the first two terms are simply \textit{link} virtualization and \textit{node} virtualization respectively, and the third is the combination of both, i.e. network virtualization. These are not ``levels'' in any meaningful sense, just different applications of virtualization. 

The term `level' (\cite{Liang2014}, \cite{Richart2016} ) is also used to refer to different levels of the protocol stack, or alternatively `depth of virtualization'/`perspective' (\cite{Wen2013}, \cite{Wen2013Current} ) is used. Some of the levels described are `flow-level virtualization', 'protocol-perspective', and 'front-end- and spectrum-perspective'. However, there is no clear distinction between any of these levels or method to recognise the virtualization level. In order to be specific and to avoid introducing ambiguity when discussing different levels of representation (i.e. virtualization levels in other works), we started this paper with a theoretical overview of AR theory, which ultimately allows us to be specific when surveying the literature.

Finally, by providing specificity, in this work we are able to derive a test for virtualization, which allows us to distinguish virtualization and non-virtualization techniques with some validity. Without specificity, the classification of works can seem arbitrary, as one does not know the difference between one class and the next.


\section{Observations and Research Questions}
\label{sect:questions}

A few conclusions can be drawn from our examination and survey of the the literature on wireless link virtualization. In this section, several observations are discussed, and we propose key open research questions regarding wireless link virtualization and wireless network virtualization.

\subsection{Comparison and Analysis of Virtualization Techniques}

\subsubsection{Metrics}
Although the benefits of wireless link/network virtualization have been discussed in an abstract manner in several papers already (\cite{Sachs2008}, \cite{Wang2013Network}, etc. ), little work has been done on measuring these benefits quantitatively. It has been very difficult to quantify the benefits of virtualization directly. Currently, to the best of our knowledge, no performance metrics exist that can measure and compare virtualized wireless networks and non-virtualized wireless networks, and show whether virtualization provides benefits in terms of isolation, flexibility, customizability, robustness, scalability and other measures, In the same vein, little work has been done which compare virtualized networks to non-virtualized networks using existing performance metrics, such as cost, revenue, efficiency, coverage, and capacity. 

\subsubsection{Requirements}
Several papers have discussed requirements for wireless link/network virtualization such as \cite{Carapinha2009}, \cite{Kalil2014}, \cite{Wen2013} and \cite{Liang2014}, but little work has been done in developing methods to compare how virtualization techniques address these requirements. Some of the requirements listed in these works are subjective and can not be measured easily such as the requirements of allowing heterogeneity and mobility and providing a generic and modular interface that is technology agnostic. To the best of our knowledge, there are currently no available methods of comparing virtualization techniques in terms of aspects discussed earlier such as isolation, flexibility, etc. 

\subsubsection{Aggregation}
Another observation is that most work is currently focused on the sharing (or slicing) of link resources. An interesting research topic would be to investigate whether virtualization can provide advantages to networks through the aggregation of resources, and what applications could benefit from this. In a sense, some existing work such as Carrier Aggregation (CA) \cite{Lee2014} could be considered to fall under this heading.  

\subsubsection{Representation Level}
A very interesting research topic could be to investigate which representation level(s) would be more suitable for virtualization for different types of wireless networks such as nomadic, mobile and ad-hoc networks. Although this question could not possibly be answered for every wireless network application, it might be possible to explore whether some applications are better suited to spectrum-level or information-level virtualization. 

One consideration to bear in mind is that due to the stochastic nature of the wireless channel, it is not possible to guarantee a specific data-rate, unlike a wired connection. This could mean that spectrum-level virtualization can offer some advantages over information-level virtualization in that it allows virtual links to have customizable spectrum-level properties, such as modulation and coding scheme, the assignment of frequencies, and the choice of technology, whereas for information-level virtualization every virtual network uses the same spectrum-level conditions and parameters. Spectrum-level virtualization could also allow spectrum resources to be used more efficiently than information-level virtualization, since wireless links can be optimized to the channel conditions and application requirements. However, the isolation required between different virtual links at the spectrum level is not as straightforward, and might require additional processing and overheads in terms of initialization and management. 

Some advantages of information-level virtualization compared to spectrum level virtualization are that it might be easier and less complex. it could also be easier to integrate with wired network virtualization, since packets with VLAN tags coming from a wired virtual network can be forwarded directly from the wireless virtual access point to the nodes. Each virtual link can also easily set its own information-level settings such as data rate, latency, security and authentication, etc. 

\subsubsection{Isolation}
Another question which has barely been addressed in the literature is the form of isolation that is most appropriate for different applications and scenarios. The only work we are aware that compares forms of isolation is \cite{Mahindra2008}. From the survey, we found that works at the spectrum-level provide isolation primarily through the frequency dimensions, and sometimes combine this with an addition dimension, while at the information-level the isolation is almost exclusively done in the time dimension.

\subsubsection{Research Questions}
Several questions emerge on these issues such as: what wireless network applications would benefit most from virtualization and which virtualization techniques are most suitable? Are there any applications that could benefit from resource aggregation? Is it possible to develop metrics to quantify virtualization techniques in terms of isolation, efficiency, flexibility, customizability, scalability, and robustness? Can we compare virtualization methods at different levels of representation using these metrics? What forms of isolation are most appropriate?

The questions shown above provide a few examples of the many questions relating to this topic, which could be investigated further and could provide many benefits to wireless networks.

\subsection{Implementation at the Spectrum Level}

Currently, the virtualization of wireless links has been implemented at the information level, and is being used commercially with success \cite{VirtualAccessPoint}. However, to the best of our knowledge, the only implementation of a link virtualization scheme at the spectrum level is the Spectrum Virtualization Layer \cite{Tan2012}. Virtualizing wireless link resources at the spectrum level could provide many advantages over the information level, since it would allow the users of virtual network resources much greater control, flexibility, and customizability. The electromagnetic spectrum is considered a very expensive and valuable resource, but it remains artificially underused in many cases \cite{Weiser2008}. An implementation at the spectrum-level would allow spectrum to be used more efficiently and tailored to specific applications, while also maintaining strict isolation between different networks.

However, to date, the majority of work on spectrum level virtualization has been theoretical in nature, and has focused on the embedding problem, that is, the issue of mapping virtual resources to real resources. The isolation problem has received very little attention, and it would be interesting to examine the different types of isolation for spectrum level virtualization, and discover the advantages and disadvantages of each. 

\subsubsection{Frameworks}
Several architectures and frameworks for spectrum level virtualization have been proposed such as \cite{Sachs2008}, \cite{Park2009}, \cite{Hoffmann2011}, and \cite{Wang2013Network}. Architectures have often been aimed at specific technologies, such as LTE \cite{Zaki2011}, and are often theoretical in nature, such as \cite{Yang2013} and \cite[Chapter 6]{Wen2013}. Although these architectures have made valuable contributions, to make further progress towards implementation, it could be necessary to consider aspects of virtualization that have received little to no attention, such as the different steps involved in virtualization, rather than only the embedding problem.

\subsubsection{Purpose of virtual networks}
Another observation that most works do not consider is the application and purpose of virtual networks and how to know how many resources they require. Only the works \cite{Zhu2015} and \cite{Zhu2015_2} consider the request strategies of virtual network users. This is a significant limitation since the users requesting virtual networks somehow need to decide how many resources to request from an infrastructure provider and which resources to request. Since virtualization allows resources to be used in a much more flexible manner than in traditional networks, conventional prediction strategies might not be suitable, and analysis should be done to investigate how virtual network users can predict how many and which resources they require, which request strategies to adopt, and also how to discover and request available resources. Different mechanisms for requesting virtual resources and virtual networks could be examined too. 

\subsubsection{Node and network virtualization}
In most of the wireless network virtualization work so far, the focus is on virtualizing the link resources and assuming that the node resources can be virtualized easily. However this might not always be the case. The current work on Network Function Virtualization tackles some of the processing issues and shows that this assumption does not apply. To the best of our knowledge the only works that consider both wireless node and link virtualization are \cite{DiStasi2013} and \cite{Rahman2016}, however for both of these works the wireless link resource allocation does not pass the recursive test, and thus the wireless link aspect of these works is not considered virtualization.

Another aspect that is not yet known to what extent is the virtualization of transmission resources possible, for example recent work on Multiple Input Multiple Output (MIMO) transmission could be considered as the aggregation of several antennas to create a virtual single directed antenna (one-to-many virtualization). It is not known yet whether it is physically possible to share one antenna between multiple virtual entities (many-to-one virtualization).

Building on an implementation of spectrum level link virtualization, it would be very interesting to apply some of the concepts developed in network virtualization and to create wireless virtual \textit{networks}, rather than virtual links only. It might be possible to create wireless network virtualization implementations that allow users to dynamically request wireless virtual networks with specific, customizable requirements.

\subsubsection{Research Questions}
A set of questions on these issues can be formed: What are the differences between isolation methods for spectrum-level virtualization and is it possible to show the advantages and disadvantages of each isolation approach?  Is it possible to develop a virtual wireless network implementation at the spectrum-level, that allows users to dynamically request and pay for virtual networks with specific node and link requirements such as locations, processing power, storage, interfacing, latency and capacity requirements? What issues arise when combining node and link virtualization for wireless network virtualization and can we apply methods from network virtualization to overcome these issues? How do virtual resource users know 1) how many and which virtual resources they need, 2) which resources are available, 3) how to request these available resources, and 4) what are the best request mechanisms and strategies to adopt?

\subsection{Discussion}

It is clear that there are still many research challenges and open questions on the topics of wireless link virtualization and wireless network virtualization. We have divided these questions into two broad categories above, questions concerning the analysis of wireless network virtualization, and questions regarding the implementation of wireless network virtualization at the spectrum level. Some of these questions might be easier to investigate, while other issues might remain unresolved. 

Although developing new metrics for comparing virtualization techniques would be a very interesting challenge, in our opinion it is more important to develop implementations at the spectrum level first. The reasons for this are that 1) Metrics for comparing virtualization techniques, either to examine which technique is most appropriate, or to see whether virtualization offers benefits over traditional networks, are only useful once these techniques exist and are implementable, 2) Implementations at the spectrum level could change the way in which spectrum resources are used, and potentially allow for new functionality and business models, 3) Implementations could aid in answering other questions, for example, issues about scalability, robustness, and flexibility, while additional problems might appear that cannot be known in advance, and finally 4) Implementation at the spectrum level would enable comparison between spectrum and information level wireless link virtualization, and analysis of suitable virtualization techniques for different applications.

\section{Conclusion}
\label{sect:conclusion}

In this work our purpose was to provide an overview of the challenges in wireless link virtualization and to consider possible next steps forward. However, before this could be accomplished, it was necessary to clarify several aspects about wireless network virtualization. We first revisited several key concepts and clarified the difference between abstraction and representation. We examined AR theory, and extended it to include the concept of levels of representation. We also showed how representation and instantiation apply to virtualization.

We developed a theory of virtualization to discuss virtualization in a coherent and structured manner. We showed that recursion is a key property of virtualization. We demonstrated how virtualization can be broken down into the isolation and the embedding problems, and provided a new definition of virtualization.

We examined network resource virtualization, which is necessary for network virtualization, starting from processing virtualization and continuing through to link virtualization. Next we described network and wireless network virtualization and showed that before WNV can be done, it is necessary to first develop wireless link virtualization. Thus we performed a survey of existing wireless link virtualization techniques by representation level and isolation method, and described several interesting observations. Finally, we observed several open problems in the area of wireless link virtualization and wireless network virtualization and formed research questions based on these problems. 

In future work we hope to examine the steps involved in virtualization and work on a spectrum-level implementation, and possibly develop new metrics for comparing virtualization techniques.

\section*{Acknowledgement}

This publication has emanated from research conducted with the financial support of Science Foundation Ireland (SFI) and is co-funded under the European Regional Development Fund under Grant Number 13/RC/2077. Special thanks to Dr. Jacek Kibilda for his insights and support.

\bibliographystyle{IEEEtran}
\bibliography{wnv_survey}

%

%
%
%




\end{document}